\documentclass[preprint,12pt]{elsarticle}

\usepackage{graphicx}

\usepackage{amssymb, amsthm}

\biboptions{compress}

\journal{Optik}

\usepackage{amsmath}
\usepackage{url}
\usepackage{color}
\usepackage{tensor}
\usepackage{mathrsfs}
\usepackage{nicefrac}
\usepackage{array}
\usepackage{ulem}
\usepackage{placeins}

\usepackage[latin1]{inputenc}      
\usepackage[T1]{fontenc}    
\usepackage{ae,aecompl}                          
\usepackage{dsfont}
\usepackage{amsfonts,amsmath,latexsym,bbm}
\usepackage{amssymb}
\usepackage{accents}
\usepackage{graphics}
\usepackage{graphicx}
\usepackage{booktabs,longtable}
\usepackage{bm}
\usepackage{dcolumn}
\usepackage{enumitem}
\usepackage{upgreek}
\usepackage{tensor}

\usepackage[linktocpage=true]{hyperref}
\setlist{font=\normalfont\itshape} 

\renewcommand{\vec}[1]{\boldsymbol{#1}}
\newcommand{\tsr}[1]{\overset\leftrightarrow{#1}}
\newcommand{\ext}{_{\rm ext}}
\newcommand{\tot}{_{\rm tot}}
\newcommand{\ind}{_{\rm ind}}

\newcommand{\h}{\hspace{1pt}}
\newcommand{\hh}{\hspace{0.5pt}}
\newcommand{\mh}{\hspace{-1pt}}
\renewcommand{\j}{\mathrm i}
\newcommand{\de}{\mathrm d}
\newcommand{\e}{\mathrm e}

\newcommand{\lar}[1]{\textnormal{\mbox{\large $#1$}}}

\newcommand{\raisemath}[1]{\mathpalette{\raisemith{#1}}}
\newcommand{\raisemith}[3]{\raisebox{#1}{$#2#3$}}

\newcommand{\EE}{_{\raisemath{-2pt}{EE}}}
\newcommand{\EB}{_{\raisemath{-2pt}{EB}}}
\newcommand{\BE}{_{\raisemath{-2pt}{BE}}}
\newcommand{\BB}{_{\raisemath{-2pt}{BB}}}

\DeclareMathAlphabet{\mathbbmsl}{U}{bbm}{m}{sl}

\numberwithin{equation}{section}

\begin{document}

\begin{frontmatter}



\title{Microscopic Theory of the Refractive Index}


\author[freiberg]{R.~Starke}
\ead{Ronald.Starke@physik.tu-freiberg.de}

\author[heidelberg]{G.A.H.~Schober\corref{cor1}}
\ead{G.Schober@thphys.uni-heidelberg.de}

\cortext[cor1]{Corresponding author.}

\address[freiberg]{Institute for Theoretical Physics, TU Bergakademie Freiberg, Leipziger Stra\ss e 23, \\ 09596 Freiberg, Germany}
\address[heidelberg]{Institute for Theoretical Physics, University of Heidelberg, Philosophenweg 19, \\ 69120 Heidelberg, Germany \vspace{-0.6cm}}

\begin{abstract}
We examine the refractive index from the viewpoint of modern first-princi\-{}ples materials physics. We first argue that the standard formula, $n^2=\varepsilon_{\mathrm r} \h\mu_{\mathrm r}\h,$ is generally in conflict with fundamental principles on the microscopic level.
Instead, it turns out that an allegedly approximate relation, \mbox{$n^2=\varepsilon_{\mathrm r}$\h,}
which is already being used for  most practical purposes, can be justified theoretically at optical wavelengths. More generally, starting from the fundamental, Lorentz-covariant electromagnetic wave equation in materials as used in plasma physics,
we rederive a well-known, three-dimensional form of the wave equation in materials and thereby clarify the connection
between the covariant {\it fundamental response tensor} and the various cartesian tensors used to describe optical properties.
Finally, we prove a general theorem by which the fundamental, covariant wave equation can be reformulated concisely in terms of the microscopic dielectric tensor. 
\end{abstract}


\end{frontmatter}



\newpage
\tableofcontents

\newpage
\section{Introduction}

By the advent of first-principles materials science \cite{Martin, Hafner08, Hafner10, Bluegel}, the past decades have witnessed an unprecedented progress in 
the quantitative description of materials properties. Typical electromagnetic response properties, such as the conductivity and the dielectric tensor, 
are now within the reach of ab initio calculations \cite{Perlov, Wang06, Baroni}, which thereby provide a new pathway to the theoretical design and optimi\-{}zation of functional materials \cite{Sato02, Shevlin, Carter, ZhangAbInitio}. 
A cornerstone in the development of first-principles electrodynamics of media has been the Modern Theory of Polarization \cite{Resta07, Resta10, Vanderbilt}, 
which first demonstrated the fallacy of simplified material models such as the Clausius-Mossotti picture of elementary electric or magnetic dipoles \cite{Mossotti}. On the other hand, 
since such simplified models had played an important conceptual r\^{o}le in the Standard Approach to electrodynamics in media \cite{Jackson, Griffiths, Landau}, the Modern 
Theory of 
Polarization also called for new perspectives on electrodynamics of materials which are based on first principles \cite{Essin2, Essin}.

Taking benefit of these ground-breaking insights, the Functional Approach to electrodynamics of materials has recently been developed by the authors of this article \cite{ED1}. 
Its aim is the axiomatization and systematic elaboration of the already existing {\it microscopic} treatments of electrodynamics in materials as developed more or less independently
in the electronic structure physics \cite{Kaxiras, Bruus, Giuliani}, semiconductor physics \cite{SchafWegener} and plasma physics \cite{Melrose1Book} communities.
The Functional Approach provides for a complete description of all linear electromagnetic materials 
properties in terms of the microscopic conductivity tensor (cf.~\cite{Smith, Dolgov, Melrose, Cho10}), a quantity which is routinely computed, for example, within the density functional theory framework \cite{VASP, Wien2k, QE-2009}. 
In particular, this approach includes the formulation of {\itshape universal response relations}, 
which are analytical formulae suitable for the ab initio computation of all linear electromagnetic response functions. 
As the Functional Approach is exclusively based on the microscopic Maxwell equations, it is independent of a~priori 
assumptions about the material, and therefore contributes to the modern pursuit of an unbiased first-principles description of \mbox{materials properties.}

The present article aims at contributing further to these developments by approaching also the optical properties of matter from first principles. 
The most important quantity in this context is of course the refractive index, by which the optical properties of many materials can be characterized \cite{Ramakrishna}. 
The refractive index is usually defined as the ratio, $n = c / u$, between the speed of light in the vacuum and in the medium. In particular, it determines via Snell's law the reflection and refraction of light at the interface between two different materials \cite{Hecht, BornWolf}. 

Ever since the foundation of classical electrodynamics by J.\,C. Maxwell in the nineteenth century, the standard formula for the refractive index,
\begin{equation} \label{eq_Maxwell_relation}
 n^2 = \varepsilon_{\rm r} \h \mu_{\rm r} \,,
\end{equation}
has been a commonplace in almost all {\it theoretical} treatments of optics and classical electrodynamics \cite{Griffiths, Landau, Hecht, BornWolf, Brooker, DiMarzio, Towne}. It relates the refractive index $n$ to the relative permittivity (or dielectric constant) $\varepsilon_{\rm r}$ and the relative permeability $\mu_{\rm r}$ of the medium. For most {\itshape practical} purposes, however, the relative permeability does not play any r\^{o}le, and in fact, it is usually {\itshape assumed} that $\mu_{\rm r} \approx 1$ would hold at optical frequencies \cite{Griffiths, Landau, Hecht, BornWolf, Brooker, DiMarzio, Towne}. Consequently, the refractive index is related only to the relative permittivity through
\begin{equation} \label{eq_Maxwell_appr}
 n^2 = \varepsilon_{\rm r} \,.
\end{equation}
This latter formula, which is often called the {\itshape Maxwell relation} \cite{Hecht}, is commonly used for extracting the frequency-dependent dielectric function from experimentally determined reflectivity spectra \cite{Lee, Laszlo, Dresselhaus, Cardona}. (Note that in the literature, also Eq.~\eqref{eq_Maxwell_relation} is sometimes called the Maxwell relation \cite{Ramakrishna}.)

Recently, the refractive index $n$ has attracted renewed interest in the field of metamaterials.
It has been argued that $n$ should be regarded as a negative number if both $\varepsilon_{\rm r} < 0$ and $\mu_{\rm r} < 0$ \cite{Veselago}. 
Concretely, a negative effective permeability occurs in artificial materials by exploiting the concept of a split ring resonator \cite{Pendry, Smith00}. An anomalous light refraction at metamaterials has been observed experimentally \cite{Shelby}. Therefore, metamaterials are regarded as promising candidates for technological applications such as superlenses \cite{PendrySuper, SmithPendry} and invisibility cloaks~\cite{Ergin}. This line of thought has been taken up in high-energy physics, where the holographic principle has been used to study the refractive index of strongly coupled gauge theories with anti-de-Sitter space duals \cite{metastring} (see also \cite{Ellis}). As the present article deals with homogeneous (or translationally invariant) systems, its results do not carry over to metamaterials, which are typically inhomogeneous, involving spatially variable response functions. However, the present article does indeed explain why a negative refractive index can, as a matter of principle, not be realized with ordinary bulk materials  as had still been hoped by V.\,G.~Veselago in his original publication \cite{Veselago}.

\pagebreak
The purpose of this article is to develop a microscopic theory of the refractive index. We will show that Eq.~\eqref{eq_Maxwell_relation} is generally in conflict with first principles, which in our case refer to the microscopic Maxwell equations and the {\itshape ab initio} calculation of microscopic response functions via the Kubo formalism. On the other hand, we will show that Eq.~\eqref{eq_Maxwell_appr}---which is already being used nowadays for 
most practical purposes (see e.g.~\cite{DiMarzio, Cardona}) and for the comparison with the experiment (see e.g.~the standard reference \cite{Hecht})---is in accord 
with fundamental field equations. More precisely, we will derive Eq.~\eqref{eq_Maxwell_appr} at optical wavelengths from the 
fundamental wave equation in materials. Independently of this wavelength restriction though, we will show that the fundamental covariant
wave equation for the electromagnetic four-potential can be concisely reformulated in terms of the microscopic dielectric tensor.
Concretely, it will turn out that the microscopic wave equation in materials simply restricts the electric field to the null-space of this microscopic 
dielectric tensor. In the case of longitudinal oscillations, where the corresponding proper oscillations are called plasmons, this is in fact well-known.
Having demonstrated that a similar logic also applies to the transverse field components, we will thus have shown that the theory of light
propagation in materials combines with the theory of plasmons into one general field equation in media.

The article is organized as follows: In Sec.~\ref{sec_FA}, we assemble the key formulae of the Functional Approach 
to electrodynamics of media as needed for this article. In Sec.~\ref{sec_SA}, we subject the Standard Approach to the refractive index to a thorough critique. In Sec.~\ref{sec_refr}, we develop the Functional Approach to the refractive index, which is based on microscopic
wave equations in materials (Subsec.~\ref{subsec_WaveEquations}). In particular, this allows for a redefinition of the speed of light in materials 
within a covariant framework (\ref{subsec_speed_light}--\ref{sec_iso}) and for a rejustification of the Maxwell relation at optical wavelengths (\ref{recon}).

\section{Functional Approach to electrodynamics of media}\label{sec_FA}

The microscopic theory of the refractive index exposed in this article is based on the Functional Approach
to electrodynamics of materials \cite{ED1,ED2,EDOhm,EffWW,EDLor}, which as a microscopic theory encapsulates the common practice in 
ab initio materials physics. For the convenience of the reader, we first assemble the most important facts
as far as they are needed in this article. The formulae will therefore neither be derived nor discussed in the following.
Instead, the interested reader is referred to Ref.~\cite{ED1} for a technical exposition of the Functional Approach or to Ref.~\cite{ED2} for a paradigmatic
discussion. 

\subsection{Field identifications}

As a matter of principle, classical electrodynamics is based on the Maxwell equations which read
\begin{align}
\nabla\cdot\vec E(\vec x,t)&=\rho(\vec x,t)/\varepsilon_0\,,\label{eq_Maxfund1}\\[3pt]
\nabla\times\vec E(\vec x,t)&=-\partial_t\vec B(\vec x,t)\,,\label{eq_Maxfund2}\\[3pt]
\nabla\cdot\vec B(\vec x,t)&=0\,,\label{eq_Maxfund3}\\[2pt]
\nabla\times\vec B(\vec x,t)&=\mu_0 \h \vec j(\vec x,t)+\varepsilon_0 \h \mu_0 \h \partial_t\vec E(\vec x,t)\,.\label{eq_Maxfund4}
\end{align}
On a fundamental, microscopic level, these equations retain their validity even within
materials (see \cite{Martin, SchafWegener, Bertlmann, Sexl, Misner, Itzykson, Fliessbach}, and in particular \cite[pp.~3~f.]{Melrose}). 
Correspondingly, in the ab initio materials physics and condensed matter physics communities, it is common practice \cite{Martin,MartinRothen,Ashcroft,Kittel}
to split the electromagnetic fields into external and induced contributions as
\begin{align}
\vec E\tot(\vec x, t)&=\vec E\ext(\vec x, t)+\vec E\ind(\vec x, t)\,,\label{eq_totEfield}\\[5pt]
\vec B\tot(\vec x, t)&=\vec B\ext(\vec x, t)+\vec B\ind(\vec x, t)\,.\label{eq_totBfield}
\end{align}
These contributions are related to the fields used in the traditional approach to electrodynamics 
in media through the {\it fundamental field identifications} (see e.g.~\cite[App.~A.2]{Kaxiras}, \cite[Sec.~6.4]{Bruus}, \cite[p.~33, footnote 14]{SchafWegener}, \cite[p.~230]{Dolgov}, \cite[Sec.~4.3.2]{MartinRothen} and \cite[p.~338, footnote 21]{Ashcroft}):
\begin{align}
 \vec P(\vec x,t) & = -\varepsilon_0\vec E\ind(\vec x, t) \,, \label{eq_PE} \\[2pt]
 \vec D(\vec x, t) & = \varepsilon_0\vec E\ext(\vec x, t) \,, \label{eq_DE} \\[2pt]
 \vec E(\vec x, t) & = \vec E\tot(\vec x, t) \,, \label{eq_EE}
\end{align}
and
\begin{align}
 \vec M(\vec x, t) & = \vec B\ind(\vec x, t)/\mu_0 \,, \label{eq_MB} \\[2pt]
 \vec H(\vec x, t) & = \vec B\ext(\vec x, t)/\mu_0 \,, \label{eq_HB} \\[2pt]
 \vec B(\vec x, t) & = \vec B\tot(\vec x, t) \,. \label{eq_BB}
\end{align}
All fields are uniquely defined by their respective Maxwell equations, which 
are given explicitly by Eqs.~\eqref{eq_Maxfund1}--\eqref{eq_Maxfund4} 
for the total fields, by
\begin{align}
\nabla\cdot\vec D(\vec x, t)&=\rho\ext(\vec x, t)\,,\label{eq_MaxDH1}\\[3pt]
\nabla\times\vec D(\vec x, t)&=-\partial_t\vec H(\vec x, t)/c^2\,,\label{eq_MaxDH2}\\[3pt]
\nabla\cdot\vec H(\vec x, t)&=0\,,\label{eq_MaxDH3}\\[3pt]
\nabla\times\vec H(\vec x, t)&=\vec j\ext(\vec x, t)+\partial_t\vec D(\vec x, t)\label{eq_MaxDH4}
\end{align}
for the external fields, and finally by
\begin{align}
\nabla\cdot\vec P(\vec x, t)&=-\rho\ind(\vec x, t)\,,\label{eq_MaxPM1}\\[3pt]
\nabla\times\vec P(\vec x, t)&=\partial_t\vec M(\vec x, t)/c^2\,,\label{eq_MaxPM2}\\[3pt]
\nabla\cdot\vec M(\vec x, t)&=0\,,\label{eq_MaxPM3}\\[3pt]
\nabla\times\vec M(\vec x, t)&=\vec j\ind(\vec x, t)-\partial_t\vec P(\vec x, t) \label{eq_MaxPM4}
\end{align}
for the induced fields. Now, the complete information about electromagnetic fields is quite generally contained in the so-called potentials, $\vec A$ and $\varphi$,
by which the fields can be represented as
\begin{align}
\vec E(\vec x,t)&=-\nabla\varphi(\vec x,t)-\partial_t\vec A(\vec x,t)\,, \label{eq_EAphi} \\[3pt]
\vec B(\vec x,t)&=\nabla\times\vec A(\vec x,t)\,.
\end{align}
In order to treat the response of a material with respect to external electromagnetic fields, it hence suffices to know the reaction to variations in the external potentials.
On the other hand, the induced electromagnetic fields are {\it per definitionem} generated by the induced current and charge densities.
The starting point of the Functional Approach is therefore the postulated functional dependence of the induced currents on the external 
potentials, as it is encapsulated in the formula \cite{Melrose1Book, Altland}
\begin{equation}
 j^\mu\ind = j^\mu\ind \h [A^\nu\ext]\,, \smallskip
\end{equation}
which relates the induced four-current $j^\mu=(c\rho, \h \vec j)^{\rm T}$ to the external four-potential $A^\nu = (\varphi/c, \h \vec A)^{\rm T}$.
In particular, {\it linear} response theory corresponds to the first-order expansion of this functional \cite{Giuliani, Altland},
\begin{equation}
j\ind^\mu(x)=\int\!\de^4 x'\,\chi\indices{^\mu_\nu}(x,x') \h A^\nu\ext(x')\,, \label{eq_fundRespRel}
\end{equation}
where $x \equiv x^\mu = (c \h t, \h \vec x)^{\rm T}$ and $\de^4 x = \de^3 \vec x \, \de x^0$. For the ubiquitous Lorentz-covariant integral kernel,
\begin{equation} \label{chimn}
\chi\indices{^\mu_\nu}(x,x') = \frac{\delta j^\mu\ind(x)}{\delta A^\nu\ext(x')}\,,
\end{equation}
we propose the name {\it fundamental response tensor}.

\subsection{Universal response relations} \label{subsec_univ}

The principles of gauge invariance and current conservation imply the following constraints on the fundamental response tensor \cite{Melrose1Book, Altland}:
\begin{align}
 \partial_\mu \h \chi\indices{^\mu_\nu}(x, x') & = 0 \,, \label{con_1} \\[5pt]
 \partial'^\nu \chi\indices{^\mu_\nu}(x, x') & = 0 \,. \label{con_2}
 \end{align}
Consequently, {\it there are at most nine independent response functions for any material} \cite{Smith, Dolgov, Melrose, Cho10}. 
In particular, the spatial (or cartesian, i.e.~$3\times 3$) {\it current response tensor},
\begin{equation}
\tsr\chi(\vec x,t;\vec x',t')=\frac{\delta\vec j\ind(\vec x,t)}{\delta\vec A\ext(\vec x',t')}\,, \smallskip
\end{equation}
determines all other response functions (such as the response of the induced electric field to the external magnetic field)
by means of so-called {\it universal response relations}.
In their most general form, these relations have been derived in \cite[Sec.~6]{ED1}.
In suitable limiting cases, the universal response relations revert to well-known response relations from the textbook literature \cite{Bruus,Giuliani}.

In the {\itshape homogeneous limit,} which is imperative for the formulation of wave equations in media (see e.g.~\cite[Eq.~(6.4)]{Toptygin}, \cite[Eq.~(2.2.9)]{Agranovich}, \cite[Part IV, Eq.~(33.15)]{Levich} and \cite[Eq.~(29.1)]{Bredov}), all response functions depend only on the difference $x - x'$ of their 
spacetime arguments, or in Fourier space on only one wave-vector $k \equiv k^\mu = (\omega/c, \h \vec k)^{\rm T}$. This limit corresponds to the idealization of a system without boundaries, i.e.~a homogeneous material filling out all of space.
The constraints \eqref{con_1}--\eqref{con_2} then imply the following expression of the fundamental response tensor in terms of 
the spatial current response tensor (cf.~\cite{Melrose1Book}):
\begin{equation}\label{generalform1}
\chi^\mu_{~\nu}(\vec k,\omega)=
\left( \!
\begin{array}{rr} -\lar{\frac{c^2}{\omega^2}} \, \vec k^{\rm T} \, \tsr{\chi}(\vec k,\omega)\, \vec k & \lar{\frac{c}{\omega}} \, \vec k^{\rm T} \, \tsr{\chi}(\vec k,\omega)\, \\
[10pt] -\lar{{\frac{c}{\omega}}} \,\h \tsr{\chi}(\vec k,\omega)\, \vec k & \, \tsr{\chi}(\vec k,\omega)\, 
\end{array} \right).
\end{equation}
From this formula, one can deduce a relativistic transformation law for the cartesian (i.e.~$3 \times 3$) current response tensor (cf.~\cite[Eq.~(30)]{EDOhm}).
By the uni\-{}versal response relations, all {\itshape physical response functions} (see \cite[Sec.~6.1]{ED1}) can be expressed in terms of this cartesian current response tensor. 
In the homogeneous case, we have the following explicit formulae:
\begin{align}
\tsr{\chi}_{EE}(\vec k,\omega) \label{eq_URR1} &= \mathbbmsl D_0(\vec k, \omega) \left( \tsr{1}\! -\! \frac{c^2 |\vec k|^2}{\omega^2}\,\tsr{P}_{\mathrm L}(\vec k)\right) \tsr\chi(\vec k,\omega)\,, \\[5pt]
\tsr\chi_{EB}(\vec k, \omega) \label{eq_URR2} &=  \mathbbmsl D_0(\vec k, \omega) \left( \tsr{1}\! -\! \frac{c^2 |\vec k|^2}{\omega^2}\,\tsr{P}_{\mathrm L}(\vec k)\right) \tsr\chi(\vec k,\omega)\left(\!-\frac{\omega}{c|\vec k|}\,\tsr R_{\mathrm T}(\vec k)\right), \\[5pt]
\tsr\chi_{BE}(\vec k, \omega) \label{eq_URR3} &=\mathbbmsl D_0(\vec k, \omega)\left(\frac{c|\vec k|}{\omega}\,\tsr R_{\mathrm T}(\vec k)\right)\tsr\chi(\vec k,\omega)\,, \\[5pt]
\tsr\chi_{BB}(\vec k,\omega) \label{eq_URR4} &=
\mathbbmsl D_0(\vec k, \omega) \left(\frac{c|\vec k|}{\omega}\,\tsr R_{\mathrm T}(\vec k)\right)\tsr\chi(\vec k,\omega)\left(-\frac{\omega}{c|\vec k|}\,\tsr R_{\mathrm T}(\vec k)\right).
\end{align}
These response functions relate the induced electric or magnetic fields to the respective external fields, i.e.~for example, $\chi_{EB}=\de\vec E\ind/\de(c\vec B\ext)$ (see \cite[Eqs.~(6.41)--(6.44)]{ED1}). Furthermore, $\mathbbmsl D_0$ denotes the Green function of the d'Alembert operator and is given in the Fourier domain by
\begin{equation} \label{gf}
\mathbbmsl D_0(\vec k, \omega) = \frac{c^2\mu_0}{-\omega^2+c^2|\vec k|^2} \,.
\end{equation}
The longitudinal and transverse projection operators $P_{\mathrm L}$ and $P_{\mathrm T}$ as well as the transverse rotation operator $R_{\mathrm T}$ are explicitly defined by their action on three-dimensional vectors as
\begin{align}
\tsr P_{\mathrm L}(\vec k) \h \vec E &= \frac{\vec k\h(\vec k\cdot\vec E)}{|\vec k|^2}\,,\\[5pt]
\tsr P_{\mathrm T}(\vec k) \h \vec E &= -\frac{\vec k\times(\vec k\times\vec E)}{|\vec k|^2} \,,\\[5pt]
\tsr R_{\mathrm T}(\vec k) \h \vec E &= \frac{\vec k\times\vec E}{|\vec k|}\,.
\end{align}
The  {\itshape (relative) permittivity} (or {\itshape dielectric tensor}) and the {\itshape (relative) magnetic \linebreak

\pagebreak \noindent
permeability} are usually defined as
\begin{align}
 \vec D & = \varepsilon_0 \h \tsr \varepsilon_{\rm r} \h \vec E \,, \\[5pt]
 \vec B & = \mu_0 \h \tsr \mu_{\rm r} \h \vec H \,.
\end{align}
By the fundamental field identifications, these standard relations translate into the microscopic definitions
\begin{align}
(\tsr\varepsilon_{\rm r})^{-1}& := \frac{\de \vec E_{\rm tot}}{\de \vec E_{\rm ext}} \,, \\[5pt]
\tsr\mu_{\rm r}& := \frac{\de \vec B_{\rm tot}}{\de \vec B_{\rm ext}} \,.
\end{align}
The permittivity and the permeability can be expressed in terms of the above response functions as
follows \cite[Sec.~6]{ED1}:
\begin{align}
(\tsr\varepsilon_{\rm r})^{-1}& = \tsr 1+\tsr\chi\EE\,, \label{permit} \\[5pt]
\tsr\mu_{\rm r}& = \tsr 1+\tsr\chi\BB\,. \label{permea}
\end{align}
As the current response tensor $\tsr \chi$ is directly accessible from {\itshape ab initio} calculations via the Kubo formalism \cite{Bruus, Giuliani, Altland, Kubo}, the universal response relations \eqref{eq_URR1}--\eqref{eq_URR4} together with the above formulae \eqref{permit}--\eqref{permea} imply that also the dielectric tensor and the magnetic permeability tensor can be calculated from first principles. In particular, this means that within an {\itshape ab initio} approach, these response functions have an independent, fixed definition which does not depend on the experiment under consideration. In the following, it has to be borne in mind that our discussion of the standard formula for the refractive index refers to these fundamental definitions of $\varepsilon_{\rm r}$ and $\mu_{\rm r}$\h.

For our discussion of the refractive index, the {\it isotropic limit} will become particularly important  (see Sec.~\ref{sec_iso}). {\it Per definitionem}, for a homogeneous and isotropic material the current response tensor has the general form
\begin{equation}
\tsr{\chi}(\vec k,\omega)=\chi_{\rm L}(\vec k,\omega) \h \tsr P_{\rm L}(\vec k)+\chi_{\rm T}(\vec k,\omega) \h \tsr P_{\rm T}(\vec k)\,,
\end{equation}
with the longitudinal and transverse response functions $\chi_{\mathrm L}$ and $\chi_{\mathrm T}$. In this case, the universal response relations simplify as follows:
\begin{align}
\tsr\chi\EE(\vec k,\omega)&=-\frac{1}{\varepsilon_0 \h \omega^2}\,\chi_{\mathrm L}(\vec k,\omega)\h\tsr P_{\mathrm L}(\vec k) \h + \h \tsr\chi\BB(\vec k,\omega)\,, \label{iso_urr1} \\[5pt]
\tsr\chi\BB(\vec k,\omega)&=\frac{c|\vec k|}{\omega}\,\h\tsr\chi\EB(\vec k,\omega)\h\tsr R_{\mathrm T}(\vec k)\,, \label{iso_urr2}  \\[5pt]
\tsr\chi\EB(\vec k,\omega)&=-\frac{\omega^2}{c^2|\vec k|^2}\,\h\tsr\chi\BE(\vec k,\omega)\,, \label{iso_urr3}  \\[5pt]
\tsr\chi\BE(\vec k,\omega)&=\frac{c|\vec k|}{\omega}\,\h\tsr\chi\BB(\vec k,\omega)\h \tsr R_{\mathrm T}(\vec k)\,, \label{iso_urr4} 
\end{align}

\vspace{3pt} \noindent
where the magnetic response function is given explicitly by
\begin{align} \label{magn_resp}
\tsr\chi\BB(\vec k,\omega)&=\mathbbmsl D_0(\vec k,\omega) \h \chi_{\mathrm T}(\vec k,\omega) \h \tsr P_{\mathrm T}(\vec k)\,.
\end{align}
In fact, in the Standard Approach to the refractive index, one considers vacuum solutions to the Maxwell equations, and these
are purely transverse ~\cite[p.~274]{Marel}. Furthermore, as will be explained below, the homogeneity and iso\-{}tropy of the material is an assumption 
which is also inherent to the Standard Approach to the refractive index. 
The above formulae clearly show that under this premise, all
response properties with respect to transverse external fields are exclusively determined by the scalar quantity $\chi_{\mathrm T}(\vec k,\omega)$. 

\subsection{Proper response functions}

The above discussion refers to {\it direct response functions}, which correspond to functional derivatives of induced fields with respect to {\it external} fields.
Equally important are the {\it proper response functions}, which correspond to functional derivatives of induced fields with respect to {\it total} fields (see e.g.~\cite{Giuliani}).
All the relations stated in the previous subsection, i.e., the constraints \eqref{con_1}--\eqref{con_2}, the general form \eqref{generalform1} in homogeneous materials, as well as the universal response relations \eqref{eq_URR1}--\eqref{eq_URR4} or \eqref{iso_urr1}--\eqref{iso_urr4} also hold for the respective proper response functions. In particular, all proper
response functions are determined by the {\itshape proper fundamental response tensor},
\begin{equation} \label{eq_fundPropRespRel}
 \widetilde \chi\indices{^\mu_\nu}(x, x')
  = \frac{\delta j^\mu\ind(x)}{\delta A^\nu\tot(x')} \,. \smallskip
\end{equation}
The latter is related to its direct counterpart, Eq.~\eqref{chimn}, through
\begin{equation}
\begin{aligned}
\chi\indices{^\mu_\nu}(x,x') & = \widetilde\chi\indices{^\mu_\nu}(x,x') \\[2pt]
 & \quad \, +\int\!\de^4 y\!\int\!\de^4 y'\,\h\widetilde\chi\indices{^\mu_\lambda}(x, y) \, (D_0)\indices{^\lambda_\rho}(y, y') \, \chi\indices{^\rho_\nu}(y',x')\,,
\end{aligned}
\end{equation}
or in a compact notation,
\begin{equation} \label{eq_Dyson}
 \chi = \widetilde \chi + \widetilde \chi \h D_0 \h \chi \,.
\end{equation}
Here, the tensorial integral kernel $D_0$ denotes the free Green function of the electromagnetic four-potential, which can be chosen in the 
Fourier domain as (see \cite{Melrose, Berestetskii} or \cite[Sec.~3.4]{ED1})
\begin{equation}\label{eq_gen_GF}
(D_0)\indices{^\mu_\nu}(k) = \mathbbmsl D_0(k)\,\delta\indices{^\mu_\nu}\,.
\end{equation}
For later purposes, we further introduce the {\it full electromagnetic Green function} by
\begin{equation} \label{full_GF}
D\indices{^\mu_\nu}(\vec x,t;\vec x',t')=\frac{\delta A^\mu\tot(\vec x,t)}{\delta j\ext^\nu(\vec x',t')}\,. \smallskip \vspace{2pt}
\end{equation}
This quantity is related to the proper fundamental response tensor by means of the Schwinger--Dyson equation \cite{Itzykson, Bjorken}
\begin{equation}
D=D_0+D_0\h \widetilde\chi \h D\,.\label{eq_Dyson1}
\end{equation}
Both \eqref{eq_Dyson} and \eqref{eq_Dyson1} can be shown by functional chain rules \cite[Sec.~5.2]{ED1}.

\subsection{Electric and magnetic solution generators}

In this subsection, we introduce the electric and magnetic solution generators $\mathbbmsl E$ and~$\mathbbmsl B$, which represent an important tool for simplifying calculations. They are defined as \cite[Sec.~4]{ED1}
\begin{align}
\tsr {\mathbbmsl E}(\vec k, \omega) & = -\varepsilon_0 \h \omega^2 \h \mathbbmsl D_0(\vec k, \omega) \left( \left(1 - \frac{c^2 |\vec k|^2}{\omega^2} \right) \tsr P_{\mathrm L} (\vec k) + \tsr P_{\mathrm T}(\vec k) \right), \label{eq_OpE} \\[5pt]
 \tsr {\mathbbmsl B}(\vec k, \omega) & = -\varepsilon_0 \h \omega^2 \h \mathbbmsl D_0(\vec k, \omega) \left( \frac{c|\vec k|}{\omega} \, \tsr R_{\mathrm T}(\vec k) \right), \label{eq_OpB}
\end{align}
and can be written more compactly as
\begin{align}
 \tsr {\mathbbmsl E}(\vec k, \omega) & = \tsr P_{\mathrm L} (\vec k) + \frac{\omega^2}{\omega^2 - c^2 |\vec k|^2} \, \tsr P_{\mathrm T}(\vec k) \,, \label{el_sol_split} \\[5pt] 
 \tsr {\mathbbmsl B}(\vec k, \omega) & = \frac{\omega \, c|\vec k|}{\omega^2 - c^2 |\vec k|^2} \, \tsr R_{\mathrm T}(\vec k) \,.
\end{align}
These dimensionless operators have a two-fold importance: On the one hand, 
the vector potential in the temporal gauge can be expressed in terms of the electric and magnetic fields by means of the {\itshape canonical functional} \cite[Eq.~(4.15)]{ED1},
\begin{align}
 \vec A(\vec k, \omega) & = \frac{1}{\j\omega} \left( \h \tsr{\mathbbmsl E}(\vec k, \omega) \, \vec E(\vec k, \omega) + \tsr{\mathbbmsl B}(\vec k, \omega) \, c \vec B(\vec k, \omega) \right). \label{eq_can_tsr}
\end{align}
On the other hand, the electric and magnetic fields can be expressed in terms of the {\itshape spatial} current by the following relations (\cite[Eqs.~(4.23)--(4.24)]{ED1}, see also \cite[Eq.~(2.9)]{Dolgov}): \vspace{-3pt}
\begin{align}
 \vec E(\vec k, \omega) & = \frac{1}{\j \omega \h \varepsilon_0} \, \tsr{\mathbbmsl E}(\vec k, \omega) \, \vec j(\vec k, \omega) \,, \label{eq_tsrE} \\[5pt]
 c \vec B(\vec k, \omega) & = \frac{1}{\j \omega \h \varepsilon_0} \, \tsr{\mathbbmsl B}(\vec k, \omega) \, \vec j(\vec k, \omega) \,. \label{eq_tsrB}
\end{align}
Therefore, the electric and magnetic solution generators can be characterized as the total functional derivatives \cite[Sec.~4.2]{ED1},
\begin{align}
 \frac{\de E_i(\vec k, \omega)}{\de j_\ell(\vec k, \omega)} & = \frac{1}{\j\omega \h \varepsilon_0} \h \mathbbmsl E_{i\ell}(\vec k, \omega) \,, \label{e_sol} \\[5pt]
 c \, \frac{\de B_i(\vec k, \omega)}{\de j_\ell(\vec k, \omega)} & = \frac{1}{\j\omega \h \varepsilon_0} \h \mathbbmsl B_{i\ell}(\vec k, \omega) \,,
\end{align}
where the total derivatives are defined as
\begin{align}
 \frac{\de E_i(\vec k, \omega)}{\de j_\ell(\vec k, \omega)} & \equiv \frac{\delta E_i(\vec k, \omega)}{\delta j_\ell(\vec k, \omega)} + \frac{\delta E_i(\vec k, \omega)}{\delta \rho(\vec k, \omega)} \h \frac{\delta \rho(\vec k, \omega)}{\delta j_\ell(\vec k, \omega)} \\[5pt]
 & = \frac{\delta E_i(\vec k, \omega)}{\delta j_\ell(\vec k, \omega)} + \frac{\delta E_i(\vec k, \omega)}{\delta \rho(\vec k, \omega)} \h \frac{k_\ell}{\omega} \,.
\end{align}
In the last step we have used the continuity equation, 
\begin{equation}
\partial_t \hh \rho + \nabla \cdot \vec j = 0\,, 
\end{equation}
which in Fourier space implies
\begin{equation}
\rho(\vec k, \omega) = \frac{\vec k \cdot \vec j(\vec k, \omega)}{\omega} \,.
\end{equation}
A demonstration of the usefulness of the electric and magnetic solution generators will be given in the next subsection.

\subsection{Conductivity relations}

Here, we assemble some useful relations between the microscopic conductivity tensor and the microscopic dielectric tensor. {\it Per definitionem}, the {\it direct conductivity}~$\sigma$ relates the induced current to the {\itshape external} electric field, i.e., in the linear regime, \smallskip
\begin{equation}
\vec j\ind=\tsr\sigma \h \vec E\ext\,. \smallskip
\end{equation}
By contrast, the {\it proper conductivity} $\widetilde\sigma$ relates the induced current to the {\itshape total} electric field, i.e.,
\begin{equation}
 \vec j\ind=\tsr{\widetilde\sigma} \h \vec E\tot\,.\label{eq_properOhmLaw} \smallskip
\end{equation}
These two quantities are interrelated by (see \cite[Eq.~(11.43)]{Kubo66} and \cite[p.~232]{Dolgov})
\begin{equation} \label{etilde}
\tsr{\widetilde\sigma}=\tsr\sigma\,\tsr{\varepsilon_{\rm r}}\,,
\end{equation}
where $\varepsilon_{\rm r}$ is the dielectric tensor. 
As every response function, the conductivity can be expressed in terms of current response tensor. 
Concretely, we have the universal response relations (cf.~\cite{Giuliani,SchafWegener,Dolgov})
\begin{align}
\tsr{\chi}(\vec x,\vec x';\omega)&={\rm i}\omega \h \tsr{\sigma}(\vec x,\vec x';\omega)\,,\label{eq_standardRel}\\
\tsr{\widetilde\chi}(\vec x,\vec x';\omega)&={\rm i}\omega \h \tsr{\widetilde\sigma}(\vec x,\vec x';\omega)\,.\label{eq_standardRel1}
\end{align}
We now derive a relation between the direct conductivity and the dielectric tensor: For this purpose, we start from the identities
\begin{equation}
 \frac{\de \vec E_{\rm tot}}{\de \vec E_{\rm ext}} \h = \h \tsr 1 + \frac{\de \vec E_{\rm ind}}{\de \vec E_{\rm ext}}
 \h = \h \tsr 1 + \frac{\de \vec E_{\rm ind}}{\de \vec j_{\rm ind}} \h \frac{\de \vec j_{\rm ind}}{\de \vec E_{\rm ext}} \,,
\end{equation}
which, using the electric solution generator \eqref{e_sol}, implies that
\begin{equation}
 (\tsr \varepsilon_{\rm r})^{-1}(\vec k, \omega) = \tsr 1 + \tsr{\mathbbmsl E}(\vec k, \omega) \, \frac{1}{\j\omega \h \varepsilon_0} \h \tsr \sigma(\vec k, \omega)\,. \label{cond_rel_1}
\end{equation}
On the other hand, we also have
\begin{equation}
 \frac{\de \vec E_{\rm ext}}{\de \vec E_{\rm tot}} \h = \h \tsr 1 - \frac{\de \vec E_{\rm ind}}{\de \vec E_{\rm tot}} \h = \h \tsr 1 - \frac{\de \vec E_{\rm ind}}{\de \vec j_{\rm ind}} \h \frac{\de \vec j_{\rm ind}}{\de \vec E_{\rm tot}} \,,
\end{equation}
and consequently,
\begin{equation}
 \tsr \varepsilon_{\rm r}(\vec k, \omega) = \tsr 1 - \tsr{\mathbbmsl E}(\vec k, \omega) \, \frac{1}{\j\omega \h \varepsilon_0} \h \tsr{\widetilde \sigma}(\vec k, \omega) \,. \label{cond_rel_2}
\end{equation}
The universal relations \eqref{cond_rel_1} and \eqref{cond_rel_2}, which can also be found in \cite[Eqs.~(2.24)--(2.25)]{Dolgov}, will prove crucial for the derivation of the refractive index in the Functional Approach.
Finally, we come to the homogeneous and isotropic limit, where
\begin{equation}
\tsr\sigma(\vec k,\omega)=\sigma_{\mathrm L}(\vec k,\omega)\tsr P_{\mathrm L}(\vec k)+\sigma_{\mathrm T}(\vec k,\omega)\tsr P_{\mathrm T}(\vec k)\,.
\end{equation}
In this limit, the universal relations \eqref{cond_rel_1} and \eqref{cond_rel_2} reduce to
\begin{equation}\label{eq_useful}
(\tsr\varepsilon_{\rm r})^{-1}(\vec k,\omega)\h =\h \tsr 1\h -\h \frac{\j\omega \h \sigma_{\rm L}(\vec k,\omega)}{\varepsilon_0 \h \omega^2} \, \tsr P_{\mathrm L}(\vec k)
\h -\h \frac{\j\omega \h \sigma_{\rm T}(\vec k,\omega)}{\varepsilon_0 \h (\omega^2-c^2|\vec k|^2)} \, \tsr P_{\mathrm T}(\vec k)\,,
\end{equation}
and respectively
\begin{equation}\label{eq_useful_proper}
\tsr{\varepsilon}_{\rm r}(\vec k,\omega) \h = \h \tsr 1 \h + \h \frac{\j\omega \h \widetilde \sigma_{\rm L}(\vec k,\omega)}{\varepsilon_0 \h \omega^2} \, \tsr P_{\mathrm L}(\vec k)
\h +\h \frac{\j\omega \h \widetilde \sigma_{\rm T}(\vec k,\omega)}{\varepsilon_0 \h (\omega^2-c^2|\vec k|^2)} \, \tsr P_{\mathrm T}(\vec k)\,.
\end{equation}
These equations can be shown most easily by using the expression \eqref{el_sol_split} for the electric solution generator.
Note that in the long-wavelength limit, the electric solution generator approaches the identity operator, i.e.,
\begin{equation}
 \lim_{|\vec k| \to 0} \tsr{\mathbbmsl E}(\vec k, \omega) = \tsr 1 \,.
\end{equation}
Hence, the above equations further simplify to
\begin{equation}
(\tsr\varepsilon_{\rm r})^{-1}(\vec k,\omega)\h \approx\h \tsr 1+\frac{1}{\j\omega \h \varepsilon_0} \h \tsr\sigma(\vec k,\omega)\,,
\end{equation}
and respectively, \smallskip
\begin{equation}
 \tsr\varepsilon_{\rm r}(\vec k, \omega)\h \approx\h \tsr 1 - \frac{1}{\j\omega \h \varepsilon_0} \h \tsr{\widetilde \sigma}(\vec k, \omega) \,. \label{wn} \smallskip
\end{equation}
These relations are commonly used in microscopic condensed matter physics and first-principles electronic structure physics (see e.g.~\cite[Eq.~(E.11)]{Martin}, \cite[Eq.~(6.51)]{Bruus} and \cite[Eq.~(1.35)]{Ashcroft}).

\section{Refractive index in the Standard Approach}\label{sec_SA}
\subsection{Standard formula for the refractive index}
\subsubsection{Standard derivation}\label{subsubsec:naive}

In its most elementary form, the {\it refractive index} is a dimensionless, real number $n$ which relates the 
so-called speed of light $u$ in a medium to the speed of light $c$ in the vacuum by means of the equation
\begin{equation}
u=\frac{c}{n}\,.\label{eq_naiveDef}
\end{equation}
According to the Standard Approach, the refractive index is related to the relative permittivity $\varepsilon_{\rm r}$
and the relative permeability $\mu_{\rm r}$ of the medium through the standard formula \cite{Griffiths, Landau, Hecht, BornWolf, Brooker, DiMarzio, Towne}
\begin{equation} \label{eq_standEqRefrInd1}
n^2=\varepsilon_{\rm r} \h\hh \mu_{\rm r}\,.
\end{equation}
The derivation of this relation in the Standard Approach can be found, for example, in \cite[Appendix A.2]{Fox} and \cite[Sec.~4.3.1]{NoltingEdyn}. This standard derivation relies on the so-called {\it macroscopic} Maxwell equations, which are usually written as
\begin{align}
\nabla \cdot \vec B(\vec x,t) & = 0\,, \label{eq_maxwell_3} \\[5pt]
\nabla \times \vec E(\vec x,t) & = -\partial_t\vec B(\vec x,t) \,, \label{eq_maxwell_4} \\[5pt]
\nabla \cdot \vec D(\vec x,t) & = \rho_{\rm f}(\vec x,t)\,, \label{eq_maxwell_1} \\[5pt]
\nabla \times \vec H(\vec x,t) & = \vec j_{\rm f}(\vec x,t) + \partial_t\vec D(\vec x,t) \,, \label{eq_maxwell_2}
\end{align}
and on the {\itshape constitutive} or {\itshape material relations} written as
\begin{align}
\vec D(\vec x,t)&=\varepsilon_0 \h\hh \varepsilon_{\rm r} \h \vec E(\vec x,t)\,,\label{eq_naiveMatRel1}\\[5pt]
\vec H(\vec x,t)&=\mu_0^{-1}\hh \mu_{\rm r}^{-1}\h \vec B(\vec x,t)\,.\label{eq_naiveMatRel2}
\end{align}
One first sets the so-called ``free'' sources to zero,
\begin{align}
\rho_{\rm f}(\vec x,t)&:=0\,, \label{eq_settozero1}\\[2pt] 
\vec j_{\rm f}(\vec x,t)&:= 0\,.\label{eq_settozero2}
\end{align}
Then the originally inhomogeneous equations \eqref{eq_maxwell_1}--\eqref{eq_maxwell_2} together with Eqs.~\eqref{eq_naiveMatRel1}--\eqref{eq_naiveMatRel2} turn into the homogeneous equations
\begin{align}
 \nabla \cdot (\varepsilon_0 \h \varepsilon_{\rm r} \h \vec E(\vec x, t)) & = 0 \label{notequiv_1} \,, \\[5pt]
 \nabla \times (\mu_0^{-1} \h \mu_{\rm r}^{-1} \h \vec B(\vec x, t)) & = \partial_t \h (\varepsilon_0 \h \varepsilon_{\rm r} \h \vec E(\vec x, t)) \,. \label{notequiv_2}
\end{align}
Now, one further {\itshape assumes} that $\varepsilon_{\rm r}$ and $\mu_{\rm r}$ are constant, such that the derivatives act only on the electric and magnetic fields. Combining the resulting equations with the homogeneous equations \eqref{eq_maxwell_3}--\eqref{eq_maxwell_4} yields
\begin{align}
\nabla\cdot\vec E(\vec x,t)&\stackrel{?}{=}0\,,\label{eq_Maxfree1}\\[5pt]
\nabla\times\vec E(\vec x,t)&=-\partial_t\vec B(\vec x,t)\,,\label{eq_Maxfree2}\\[5pt]
\nabla\cdot\vec B(\vec x,t)&=0\,,\label{eq_Maxfree3}\\[2pt]
\nabla\times\vec B(\vec x,t)&\stackrel{?}{=} \varepsilon_0\h \varepsilon_{\rm r} \, \mu_0 \h \mu_{\rm r} \,\partial_t\vec E(\vec x,t)\,.\label{eq_Maxfree4}
\end{align}
In particular, Eqs.~\eqref{eq_maxwell_1} and \eqref{eq_settozero1} together with Eq.~\eqref{eq_Maxfree1} imply
that the fields in Eq.~\eqref{eq_naiveMatRel1} are purely transverse, and hence the involved material constant $\varepsilon_{\rm r}$
has in this context to be interpreted as the {\it transverse dielectric function} $\varepsilon_{\rm r,T}$\hh.
Anyway, with the standard vector identity
\begin{equation}
\nabla(\nabla\cdot\vec A)-\nabla\times(\nabla\times\vec A)=\Delta\vec A\,,
\end{equation}
one shows that Eqs.~\eqref{eq_Maxfree1}--\eqref{eq_Maxfree4} imply the relations
\begin{align}
\left(\varepsilon_0\h\varepsilon_{\rm r} \, \mu_0\h\mu_{\rm r}\,\frac{\partial^2}{\partial t^2}-\Delta\right) \mh \vec E(\vec x,t)&\stackrel{?}{=}0\,,\label{eq_freeWaveEq1}\\[3pt]
\left(\varepsilon_0\h\varepsilon_{\rm r} \, \mu_0\h\mu_{\rm r}\, \frac{\partial^2}{\partial t^2}-\Delta\right) \mh \vec B(\vec x,t)&\stackrel{?}{=}0\,.\label{eq_freeWaveEq2}
\end{align}
By a Fourier transformation, these are equivalent to
\begin{align}
\left(-\frac{\omega^2}{c^2} \h \varepsilon_{\rm r} \h \mu_{\rm r}+|\vec k|^2\right)\mh \vec E(\vec k,\omega)&\stackrel{?}{=}0\,,\label{eq_freeWaveMediaFourierEq1}\\[3pt]
\left(-\frac{\omega^2}{c^2}\h \varepsilon_{\rm r}\h \mu_{\rm r}+|\vec k|^2\right) \mh \vec B(\vec k,\omega)&\stackrel{?}{=}0\,,\label{eq_freeWaveMediaFourierEq2}
\end{align}
where $c = 1/\sqrt{\varepsilon_0 \h \mu_0}$ \,is the speed of light in the vacuum. Componentwise, the above equations \eqref{eq_freeWaveEq1}--\eqref{eq_freeWaveEq2} are equivalent to the standard wave equation,
\begin{equation} \label{original_definition}
\left(\frac{1}{u^2}\h \frac{\partial^2}{\partial t^2}-\Delta\right)\varphi(\vec x,t)=0\,,
\end{equation}
provided that one identifies the wave-velocity $u$ with
\begin{equation} \label{res_u}
u\stackrel{?}{=}\frac{c}{\sqrt{\varepsilon_{\rm r} \h \mu_{\rm r}}}\,.
\end{equation}
A direct comparison with Eq.~\eqref{eq_naiveDef} yields the desired relation \eqref{eq_standEqRefrInd1}.

We will show below that the equations \eqref{eq_Maxfree1} and \eqref{eq_Maxfree4}, the ensuing 
wave equations \eqref{eq_freeWaveEq1}--\eqref{eq_freeWaveEq2}, as well as the result \eqref{res_u} and the standard formula for the refractive index \eqref{eq_standEqRefrInd1} are all untenable. 
Instead, it will turn out that the allegedly approximate wave equations (see e.g.~\cite[Eq.~(2.203)]{SchafWegener}, \cite[Eq.~(1.34)]{Ashcroft} and \cite[Eq.~(16.18)]{Platzmann}) 
\begin{align}
\left(-\frac{\omega^2}{c^2} \h \varepsilon_{\rm r}(\vec k, \omega)+|\vec k|^2\right) \mh \vec E(\vec k,\omega)&=0\,,\label{eq_certainJust1}\\[5pt]
\left(-\frac{\omega^2}{c^2} \h \varepsilon_{\rm r}(\vec k, \omega)+|\vec k|^2\right) \mh \vec B(\vec k,\omega)&=0\,,\label{eq_certainJust2}
\end{align}
can be justified in the r\'{e}gime of optical wavelengths. These equations correspond to the so-called {\itshape Maxwell relation} \cite{Griffiths, Landau, Hecht, BornWolf, Brooker, DiMarzio, Towne},
\begin{equation}
 n^2 = \varepsilon_{\rm r} \,,
\end{equation}
which will hence turn out to be more fundamental than its allegedly exact version \eqref{eq_standEqRefrInd1}.  Furthermore, we will show in Sec.~\ref{sec_contradiction} that for a homogeneous and isotropic material, Faraday's law \eqref{eq_Maxfund2} implies the identity
\begin{equation}
 \varepsilon_{\rm r, \hh T}(\vec k, \omega) \h\hh \mu_{\rm r}(\vec k, \omega) \equiv 1 \,,
\end{equation}
where $\varepsilon_{\rm r, \hh T}(\vec k, \omega)$ denotes the transverse dielectric function (which is the relevant response function in optical experiments, cf.~\cite[p.~274]{Marel}). This shows that the standard formula \eqref{eq_standEqRefrInd1} cannot be true, because it would imply the refractive index to be always identical to one.

\subsubsection{R\^{o}le of the permeability}

While many textbooks derive the standard formula $n^2 = \varepsilon_{\rm r} \h \mu_{\rm r}$
as a fun-\linebreak damental relation \cite{Griffiths, Landau, Hecht, BornWolf, Brooker, DiMarzio, Towne, Fox}, it is actually only true 
that the relation $n^2\approx\varepsilon_{\rm r}$ (here referred to as Maxwell relation, cf.~\cite{Hecht}) is an approximation 
which holds at optical wavelengths. Below, we will provide theoretical evidence for this and show that 
the standard formula is generally in conflict with fundamental  principles on the microscopic level. Besides this, 
it is of course also crucial to provide for a comparison with the experiment.

In fact, while both the permittivity $\varepsilon_{\rm r}$ and the permeability $\mu_{\rm r}$ enter symmetrically into the standard formula, only the permittivity is usually considered for most practical purposes and for the comparison with the experiment. As mentioned in the introduction, it is 
usually {\itshape assumed} that $\mu_{\rm r} \approx 1$ would hold in all materials at optical frequencies 
(see e.g.~the standard references \cite[Eq.~(9.70)]{Griffiths}, \cite[\S~79]{Landau}, \cite[Eq.~(3.61)]{Hecht}, \cite[Problem 1.4]{Brooker}, \cite[Eq.~(1.9)]{DiMarzio}, \cite[Eq.~(6--15)]{Towne} and \cite[Eq.~(A.31)]{Fox}).
In particular, the famous textbook by Landau and Lifshitz \cite[\S~79]{Landau} even claims that the very notion of a magnetic susceptibility would loose its meaning
at optical frequencies (see also \cite{Agranovich09, Merlin}). 
Interestingly, the recipe that $\mu_{\rm r}$ should be set to one in the standard formula can already be found in the older literature. For example, in the optics treatise \cite{Gehrcke} (published in 1926) 
it is stated explicitly (p.~670) that even for ferromagnetic materials one has to set $\mu_{\rm r} = 1$ 
(which, of course, raises the question of when the magnetic permeability would ever become relevant in the standard formula). Furthermore, it is stated there (p.~730)
that the standard formula holds only in the limit of infinitely long waves (i.e., for $|\vec k| \to 0$).

We also note that in semiconductor physics and condensed matter physics, one often derives directly the approximate wave equations \eqref{eq_certainJust1}--\eqref{eq_certainJust2} 
without the detour given by Eqs.~\eqref{eq_freeWaveMediaFourierEq1}--\eqref{eq_freeWaveMediaFourierEq2} (see e.g.~\cite{SchafWegener, Ashcroft, Platzmann}).
The standard reference by P.\,Y.~Yu and M.~Cardona on semiconductor physics directly introduces the frequency-dependent refractive index by \cite[Eq.~(6.11)]{Cardona}
\begin{equation} \label{sta}
 n(\omega) = \sqrt{\varepsilon_{\rm r}(\omega)} \,.
\end{equation}
The same applies to the standard textbooks in solid state physics by N.\,W. Ashcroft and N.\,D.~Mermin \cite[p.~534]{Ashcroft} and by C.~Kittel \cite[Chap.~11, Eq.~(3)]{Kittel}. 
Furthermore, the relation \eqref{sta} combined with a Kramers-Kronig analysis forms the basis for the 
experimental determination of the dielectric function and therefore also the optical conductivity from  measured reflectivity spectra (see e.g.~\cite{Lee, Laszlo, Dresselhaus} and \cite[Chap.~6]{Cardona}).

As we will show below, the preconception that $\mu_{\rm r} \approx 1$ would generally \linebreak hold at optical frequencies, is actually not true. Quite to the contrary, 
it follows directly from the Maxwell equations that in the homogeneous and isotropic limit, 
the permeability always equals the inverse of the transverse permittivity (see Sec.~\ref{sec_contradiction}). The fact that 
in the majority of experimental works the permeability in the standard formula is set to one, therefore matches our theoretical evidence 
that \h$n^2 = \varepsilon_{\rm r}$ \h is the more correct formula for the refractive index (being valid at optical wavelengths, see Sec.~\ref{recon}). 

Before we come to a systematic investigation of the refractive index in the Functional Approach to electrodynamics of media, we first assemble the problems of the standard formula and its derivation.

\subsection{Problems of the standard formula} \label{sec_refute}

In this subsection, we systematically develop our criticism of the standard formula for the refractive index. Concretely, we rest our case on four main arguments to be spelled out below: (i) In materials, the electromagnetic wave equations always have to be inhomogeneous. (ii) The standard derivation is inconsistent, because it first treats the response functions as constants but subsequently re-introduces their frequency dependences. 
(iii) On conceptual grounds, in the homogeneous and isotropic limit any material should be described at optical frequencies by only one transverse response function. (iv) The standard formula is in conflict with the Maxwell equations, or more specifically, with Faraday's law.

\subsubsection{Inhomogeneous wave equations} \label{inhomo}

As a matter of principle, {\it induced fields cannot be vacuum fields}, because they are generated by the induced charge and current densities.
Correspondingly, instead of free wave equations one should use the fundamental, inhomogeneous wave equations for the electric and magnetic fields 
in terms of their sources (cf.~\cite[Eqs.~(6.49)--(6.50)]{Jackson} and \cite[Eqs.~(2.51)--(2.52)]{ED1}).
Applying these equations to the induced quantities yields
\begin{align}
\left(\frac{1}{c^2}\frac{\partial^2}{\partial t^2}-\Delta\right) \mh \vec E\ind(\vec x,t)&=-\frac{1}{\varepsilon_0} \h \nabla\rho\ind(\vec x,t) - \mu_0 \h \frac{\partial}{\partial t} \, \vec j\ind(\vec x,t)\,,\label{eq_sourceWave1}\\[3pt]
\left(\frac{1}{c^2}\frac{\partial^2}{\partial t^2}-\Delta\right) \mh \vec B\ind(\vec x,t)&=\mu_0 \h \nabla\times\vec j\ind(\vec x,t)\,.\label{eq_sourceWave2}
\end{align}
As such, these inhomogeneous equations are far from being free (i.e.~homo\-{}geneous) wave equations with a modified wave-velocity. In Sec.~\ref{sec_refr}, 
we will discuss in detail the question under which conditions Eqs.~\eqref{eq_sourceWave1}--\eqref{eq_sourceWave2} still take the form of a wave equation. For the purposes of this subsection, we only stress that for generating induced fields, the presence of induced sources is necessary.

We will now calculate these induced sources within the Standard Approach (treating $\varepsilon_{\rm r}$ and $\mu_{\rm r}$ as constants as in Eqs.~\eqref{eq_naiveMatRel1}--\eqref{eq_naiveMatRel2}), and show that this leads to a conflict with a well-established result of first-principles materials physics. First, we consider the charge density: By the assumption that the external sources vanish, we have $\rho_{\rm ext} = 0$ and
$\rho_{\rm tot} \equiv \rho_{\rm ext} + \rho_{\rm ind} = \rho_{\rm ind}$\h.
Using Gauss' law \eqref{eq_Maxfund1}, this implies on the one hand
\begin{equation}
 0 \h = \h \nabla \cdot \vec E_{\rm ext} \h = \h \nabla \cdot (\varepsilon_{\rm r} \h \vec E_{\rm tot}) \h \stackrel{?}{=} \h \varepsilon_{\rm r} \h \nabla \cdot \vec E_{\rm tot} \,,
\end{equation}
and on the other hand
\begin{equation}
 \rho_{\rm ind}/\varepsilon_0 \h = \h \rho_{\rm tot}/\varepsilon_0 \h = \h \nabla \cdot \vec E_{\rm tot} \,.
\end{equation}
Together, these equations imply that the product of $\varepsilon_{\rm r}$ and $\rho_{\rm ind}$ is zero, and consequently,  for $\varepsilon_{\rm r} \not = 0$\h,
\begin{equation}
\rho_{\rm ind} \stackrel{?}{=} 0 \,. \smallskip \vspace{2pt}
\end{equation}
Hence, there is only an induced {\itshape spatial} current in the Standard Approach, for which we now also derive an explicit expression: Using that $\vec j_{\rm ext} = 0$, Amp\`{e}re's law \eqref{eq_Maxfund4} for the external fields simply reads
\begin{equation}
 \nabla \times \vec B\ext(\vec x,t)= \varepsilon_0 \h \mu_0 \, \partial_t \vec E\ext(\vec x,t)\,.
\end{equation}
From Eqs.~\eqref{eq_naiveMatRel1}--\eqref{eq_naiveMatRel2} it then follows that
\begin{equation}
 \nabla \times \vec B\tot \stackrel{?}{=} \varepsilon_0 \h \varepsilon_{\rm r} \, \mu_0 \h \mu_{\rm r} \, \partial_t \vec E\tot \smallskip\,.
\end{equation}
On the other hand, the total fields have to obey the Maxwell equations as well, whence it follows that
\begin{align}
 \nabla \times \vec B\tot(\vec x,t) & =\mu_0 \h \vec j\ind(\vec x,t)+\varepsilon_0\h  \mu_0\, \partial_t\vec E\tot(\vec x,t)\,,
 \end{align}
where we have used the equality \h$\vec j\tot \equiv\vec j\ext+\vec j\ind=\vec j\ind$\h.
By comparing these two equations, we obtain the expression
\begin{equation} \label{eq_odd}
\vec j\ind(\vec x,t) \stackrel{?}{=} \varepsilon_0  \h (\varepsilon_{\rm r}\h\mu_{\rm r}-1) \, \partial_t\vec E\tot(\vec x,t) \,.
\end{equation}
This is the desired formula for the induced current, which has been derived in the Standard Approach to electrodynamics in media by treating the response functions $\varepsilon_{\rm r}$ and $\mu_{\rm r}$ as if they were constants.

It remains to show that Eq.~\eqref{eq_odd} is at odds with a standard result from first-principles materials physics. For that purpose, we first reformulate this equation in the Fourier domain as
\begin{equation}
\vec j\ind(\vec k,\omega)\stackrel{?}{=}\varepsilon_0 \h (1 - \varepsilon_{\rm r} \h \mu_{\rm r}) \, \j\omega \vec E\tot(\vec k,\omega)\,.
\end{equation}
Now, a linear relation between the induced current and the total electric field is {\it per definitionem} given by
the proper conductivity:
\begin{equation}
\vec j\ind(\vec k,\omega)=\widetilde\sigma(\vec k,\omega) \h \vec E(\vec k,\omega)\,.
\end{equation}
Comparison of these two equations yields
\begin{equation}
\varepsilon_{\rm r} \h \mu_{\rm r}\overset{?}{=}1-\frac{1}{\j\omega \h \varepsilon_0} \h \widetilde\sigma(\vec k,\omega) \,.\label{eq_atodds}
\end{equation}
This result contradicts the relation \eqref{wn}, which is well-established, for instance, in ab initio electronic structure theory. Note, however,
that the correct relation \eqref{wn} is recovered from \eqref{eq_atodds} if we substitute $\varepsilon_{\rm r} \h \mu_{\rm r}\mapsto\varepsilon_{\rm r}$\h.

\subsubsection{Frequency-dependent material constants}

In general and as a matter of principle, the material ``constants'' $\varepsilon_{\rm r}$ and $\mu_{\rm r}$ are actually frequency-dependent functions, i.e.,
\begin{align}
\varepsilon_{\rm r}&\equiv\varepsilon_{\rm r}(\omega)\,,\\[3pt]
\mu_{\rm r}&\equiv\mu_{\rm r}(\omega)\,.
\end{align}
This is necessary for the following reasons:
\begin{enumerate}
\item[(i)] {\itshape Experimental evidence.} The frequency dependence of refraction (i.e.~the {\itshape dispersion}) is an experimental fact (see e.g.~\cite[Sec.~7.5]{Jackson} and \cite[Sec.~3.5.1]{Hecht}).
In particular, it is well-known that by inserting the {\itshape static} values of $\varepsilon_{\rm r}$ and $\mu_{\rm r}$ into the standard formula \eqref{eq_standEqRefrInd1}, one may obtain completely wrong results for the index of refraction (e.g.~in the case of  water, see~\cite[Table 3.2]{Hecht}).
\item[(ii)] {\itshape Transverse light waves.} Theoretically, it makes no sense to identify~$\varepsilon_{\rm r}$ generally
with the static dielectric constant. A frequency-independent response function relates {\it static}
electric fields, and the latter are {\it longitudinal}. By contrast, light waves involve transverse electric fields, 
and hence their effects in a material should be described by transverse response functions (cf.~\cite[p.~274]{Marel}).
\end{enumerate}

\medskip \noindent
Now, the frequency-dependent response functions are of course none other than the Fourier transforms of their counterparts in the time-domain (cf.~\cite[Sec.~2.1]{ED1}), i.e.,
\begin{align}
\varepsilon_{\rm r}(\omega) & = c \int_{-\infty}^\infty\!\de\tau\,\varepsilon_{\rm r}(\tau) \,\e^{{\rm i}\omega\tau}\,,\\[6pt]
\mu_{\rm r}(\omega) & = c \int_{-\infty}^\infty\!\de\tau\,\mu_{\rm r}(\tau)\,\e^{{\rm i}\omega\tau} \,,
\end{align}
where $\tau=t-t'$ denotes the difference between the two time arguments. The standard derivation of the refractive index, which treats the response functions as constants, 
can therefore not be upheld, because any multiplication with a material constant actually involves a {\it temporal convolution}. Concretely, when $\varepsilon_{\rm r}$ and $\mu_{\rm r}$ become integral kernels, the equations \eqref{eq_freeWaveEq1}--\eqref{eq_freeWaveEq2} 
cease to be standard wave equations of the form \eqref{original_definition} and, in particular, Eq.~\eqref{res_u} looses its meaning.

Furthermore, the frequency dependence of the response functions is only 
the tip of the iceberg, because the material relations \eqref{eq_naiveMatRel1}--\eqref{eq_naiveMatRel2} are 
to be interpreted on the microscopic scale as (cf.~\cite[Eqs.~(2.16)--(2.17)]{Dolgov})
\begin{align}
\vec D(\vec x,t)&=\varepsilon_0\int\!\de^3\vec x'\!\int\!c\,\de t'\,\h\tsr\varepsilon_{\rm r}(\vec x,\vec x';t-t') \, \vec E(\vec x',t')\,,\\[3pt]
\vec H(\vec x,t)&=\mu_0^{-1}\int\!\de^3\vec x'\! \int\!c\,\de t'\,\h(\tsr\mu_{\rm r})^{-1}(\vec x,\vec x';t-t') \, \vec B(\vec x',t')\,.
\end{align}
This means that in the most general case, for example, Eq.~\eqref{eq_Maxfree4} should be replaced by
\begin{equation} \label{eq_general}
\begin{aligned}
 & \nabla \times \left( \mu_0^{-1}\int\!\de^3\vec x'\! \int\!c\,\de t'\,\h(\tsr\mu_{\rm r})^{-1}(\vec x,\vec x';t-t') \, \vec B(\vec x',t') \right) \\[5pt]
 & - \partial_t \left( \varepsilon_0\int\!\de^3\vec x'\! \int\!c\,\de t'\,\h\tsr\varepsilon_{\rm r}(\vec x,\vec x';t-t') \, \vec E(\vec x',t') \right) = \vec j_{\rm ext}(\vec x, t) \,.
\end{aligned}
\end{equation}
With these relations, however, it is impossible to derive a wave equation in media by the standard procedure.
In fact, with $\varepsilon_{\rm r}$ and $\mu_{\rm r}$ being tensorial integral kernels depending on two spacetime arguments, it is a~priori not clear what the standard formula \eqref{eq_standEqRefrInd1} for the refractive index even means.

In fairness to the Standard Approach, however, we have to concede that in order to recover a wave equation for the propagation of light in materials, one has to employ a number of appropriate approximations (cf.~\cite[Sec.~1.3]{Brooker}):
First, one has to restrict attention to homogeneous systems, where the response functions depend only on the differences of their space and time arguments, or in the Fourier domain on only one wave-vector (cf.~\cite[Eq.~(6.4)]{Toptygin}, \cite[Eq.~(2.2.9)]{Agranovich}, \cite[Part IV, Eq.~(33.15)]{Levich} and \cite[Eq.~(29.1)]{Bredov}). 
Furthermore, one has to assume the isotropy of the material, such that the transverse response functions simply read
\begin{align}
\tsr\varepsilon_{\rm r}(\vec k,\omega)&=\varepsilon_{\rm r, \hh T}(\vec k,\omega) \, \tsr P_{\rm T}(\vec k)\,,\\[2pt]
\tsr\mu_{\rm r}(\vec k,\omega)&=\mu_{\rm r}(\vec k,\omega) \, \tsr P_{\rm T}(\vec k)\,.
\end{align}
In this case, Eq.~\eqref{eq_general} first simplifies~to
\begin{equation}
\begin{aligned}
 & \mu_0^{-1}\int\!\de^3\vec x'\! \int\!c\,\de t'\,\h(\tsr\mu_{\rm r})^{-1}(\vec x - \vec x';t-t') \,\h (\nabla \times \vec B)(\vec x',t')  \\[2pt]
 & - \varepsilon_0\int\!\de^3\vec x'\! \int\!c\,\de t'\,\h\tsr\varepsilon_{\rm r}(\vec x - \vec x';t-t') \,\h (\partial_t \vec E)(\vec x',t') = \vec j_{\rm ext}(\vec x, t) \,,
\end{aligned}
\end{equation}
and hence, in the Fourier domain,
\begin{equation}
\begin{aligned}
 & \mu_0^{-1} \h ( \tsr\mu_{\rm r})^{-1}(\vec k, \omega) \ \j \vec k \times \vec B(\vec k, \omega) - \varepsilon_0 \h \tsr\varepsilon_{\rm r}(\vec k, \omega) \, (-\mathrm i\omega) \h \vec E(\vec k, \omega) = \vec j_{\rm ext}(\vec k, \omega) \,.
\end{aligned}
\end{equation}
With these relations, the standard equations \eqref{eq_freeWaveMediaFourierEq1}--\eqref{eq_freeWaveMediaFourierEq2} in Fourier space finally generalize to
\begin{align}
\left(-\frac{\omega^2}{c^2} \, \varepsilon_{\rm r, \hh T}(\vec k,\omega) \h \mu_{\rm r}(\vec k,\omega)+|\vec k|^2\right) \mh \vec E(\vec k,\omega)&\overset{?}{=}0\,, \label{eq_WaveFourierEq1} \\[3pt]
\left(-\frac{\omega^2}{c^2} \, \varepsilon_{\rm r, \hh T}(\vec k,\omega) \h \mu_{\rm r}(\vec k,\omega)+|\vec k|^2\right) \mh \vec B(\vec k,\omega)&\overset{?}{=}0\,. 
\end{align}
By an inverse Fourier transformation, however, these equations would still not revert to wave equations of the form \eqref{eq_freeWaveEq1}--\eqref{eq_freeWaveEq2}, because any product in the Fourier domain corresponds to a convolution in real space.

\subsubsection{Redundancy of transverse response functions}

The above considerations have shown that the homogeneous and isotro\-{}pic limit is inherent to the Standard 
Approach to the refractive index. However, it follows from the universal response relations that in this limit, the transverse electromagnetic response is completely described by a single response function. For example, all transverse electromagnetic response functions can be expressed in terms of the transverse current response function~$\chi_{\mathrm T}$\h. From Eqs.~\eqref{iso_urr1} and \eqref{magn_resp} we obtain in particular the relations
\begin{equation} \label{uni_id}
\chi_{EE, \h {\rm T}}(\vec k, \omega) = \chi_{BB, \h {\rm T}}(\vec k, \omega) = \mathbbmsl D_0(\vec k, \omega) \h \chi_{\rm T}(\vec k, \omega) \,.
\end{equation}
This raises doubts about the standard formula $n^2=\varepsilon_{\rm r}\h\mu_{\rm r}$\h, which apparently
implies that the optical behaviour is described by two independent response functions. By contrast,
in microscopic treatises (such as \cite{Agranovich,KeldyshKirzhnitz}), it is made clear that this not the case: For example, D.\,A.~Kirzhnitz writes that ``the quantities $\varepsilon_{\rm r, T}$ and $\mu_{\rm r}$ have no independent
meaning'' (\cite[Chap.~2, p.~47]{KeldyshKirzhnitz}, notation adapted) and later even uses this arbitrariness
to equate the squared refractive index with the transverse dielectric function (see p.~62).
Similarly, V.\,M.~Agranovich and V.\,L.~Ginzburg explicitly state that ``the tensor $\varepsilon_{ij}(\omega, \vec k)$
completely describes both the electrical and the magnetic properties of the medium'' \cite[p.~23]{Agranovich}.

\subsubsection{Contradiction with Faraday's law} \label{sec_contradiction}

Finally, we show that the standard formula \eqref{eq_standEqRefrInd1} cannot be true as a matter of principle,
not even after the re-identification of $\varepsilon_{\rm r}$ and $\mu_{\rm r}$ with frequency and wave-vector dependent transverse response functions.

Being one of the four fundamental Maxwell equations, Faraday's law \eqref{eq_Maxfund2} is valid for all electromagnetic fields.
It allows to convert {\itshape purely transverse electric and magnetic fields}---which in fact describe light waves---into each other by means of (see \cite[Eqs.~(4.25)--(4.26)]{ED1})
\begin{align}
\vec B(\vec k,\omega)&=\frac{\vec k}{\omega}\times\vec E(\vec k,\omega)\,,\label{eq_BtoE}\\[5pt]
\vec E(\vec k,\omega)&=-\frac{\omega}{|\vec k|^2} \, \vec k\times\vec B(\vec k,\omega)\,.\label{eq_EtoB}
\end{align}
Now, assuming for the transverse electric fields the relation
\begin{equation}
\vec E\ext(\vec k,\omega)=\varepsilon_{\rm r, \hh T}(\vec k,\omega) \h \vec E\tot(\vec k,\omega) \,,
\end{equation}
implies with Eqs.~\eqref{eq_BtoE}--\eqref{eq_EtoB} the analogous relation between the respective magnetic fields, i.e.,
\begin{equation}
\vec B\ext(\vec k,\omega)=\varepsilon_{\rm r, \hh T}(\vec k,\omega) \h \vec B\tot(\vec k,\omega)\,. 
\end{equation}
Comparing this result with the defining equation for the permeability,
\begin{equation}
 \vec B_{\rm tot}(\vec k, \omega) = \mu_{\rm r}(\vec k, \omega) \h \vec B_{\rm ext}(\vec k, \omega) \,,
\end{equation}
yields the relation between the two transverse response functions,
\begin{equation}
\mu_{\rm r}^{-1}(\vec k,\omega)=\varepsilon_{\rm r, \hh T}(\vec k,\omega)\,.
\end{equation}
Thus, we have shown that Faraday's law implies the identity
\begin{equation}
 \varepsilon_{\rm r, \hh T}(\vec k,\omega) \, \mu_{\rm r}(\vec k, \omega) \equiv 1 \,. \label{idfar}
\end{equation}
Now, if the standard formula \eqref{eq_standEqRefrInd1} was true, then the refractive index would always be identical to one; since this is counterfactual, we have to refuse the standard formula.

We remark that the fundamental relation \eqref{idfar} can be derived independently from the universal response relations: the equation \eqref{uni_id} together with \eqref{permit}--\eqref{permea} implies immediately that
\begin{equation} \label{surpr}
 \varepsilon_{\rm r, T}^{-1}(\vec k, \omega) = \mu_{\rm r}(\vec k, \omega) = 1 + \mathbbmsl D_0(\vec k, \omega) \h \chi_{\rm T}(\vec k, \omega) \,.
\end{equation}
The equality of the permeability and the transverse part of the inverse dielectric function
in the homogeneous and isotropic limit, also shows that it is generally {\itshape not true} that \mbox{$\mu_{\rm r} \approx 1$} holds at optical frequencies. Finally, we note that, although the equality \eqref{surpr} may come as a complete surprise from the point of view of the Standard Approach to electrodynamics in media, it is actually not new in condensed matter physics.
For example, L.\,V. Keldysh clearly states that ``in any system, because of the unique correspondence between the alternating magnetic field and the solenoidal electric field,
the response to an arbitrary magnetic field may be regarded as the response to the attendant solenoidal (transverse) electric field or it may be divided arbitrarily
into two parts, one of which is considered to be the response to the magnetic field and the other as the response to the 
transverse electric field'' \cite[Chap.~1, p.~8]{KeldyshKirzhnitz}.

\section{Refractive index in the Functional Approach}\label{sec_refr}

In the previous section we have shown that the standard wave equations \eqref{eq_freeWaveEq1}--\eqref{eq_freeWaveEq2} 
and the ensuing formula for the refractive index cannot be upheld. We therefore face two fundamental questions:
\begin{enumerate}
 \item[(i)] How do the fundamental, microscopic, inhomogeneous wave equations \eqref{eq_sourceWave1}--\eqref{eq_sourceWave2}
 revert to the form of a homogeneous wave equation with a modified speed of light?
 \item[(ii)] How is that modified speed of light determined?
\end{enumerate}
Based on the Functional Approach to electrodynamics of media, we will investigate in this section microscopic
wave equations in materials, which only assume the linearity and homogeneity of the medium, but
incorporate all effects of non-locality, relativistic retardation, anisotropy and magnetoelectric cross-coupling. 
From these, we will subsequently find the answers to the questions {\itshape (i)} and {\itshape (ii)} in Secs.~\ref{subsec_WaveEquations} and \ref{subsec_speed_light}, respectively.

\subsection{Wave equations in materials}\label{subsec_WaveEquations}
\subsubsection{Fundamental covariant equation}

The fundamental equation of motion for the electromagnetic four-potential in terms of {\it its} generating four-current reads
\begin{equation} \label{fund_eom}
(\eta\indices{^\mu_\nu}\Box+\partial^\mu\partial_\nu) \h A^\nu=\mu_0 \h j^\mu\,,
\end{equation}
where 
\begin{equation}
\Box = -\partial^\mu \partial_\mu =\frac{1}{c^2}\frac{\partial^2}{\partial t^2}-\Delta \smallskip \vspace{2pt}
\end{equation}
is the d'Alembert operator. As the Functional Approach is inherently microscopic, 
all fundamental field equations (i.e.~the Maxwell equations) and consequently also Eq.~\eqref{fund_eom}
carry over to the induced fields $A^\nu\ind$ and $j^\mu\ind$\hh.\linebreak
Furthermore, as in the Standard Approach we consider the case of vanishing external sources, such that the total four-potential $A^\mu \equiv A^\mu_{\rm tot}$ obeys the inhomogeneous wave equation with the induced sources, i.e.,
\begin{equation}
(\eta\indices{^\mu_\nu}\Box+\partial^\mu\partial_\nu) \h A^\nu=\mu_0 \h j^\mu\ind \,.
\end{equation}
In order to re-interpret this {\it inhomogeneous}
wave equation as a {\it modified homogeneous} (or free) wave equation, we have to eliminate the induced four-
current in favor of
the total four-potential. For this purpose, we use a standard procedure (cf.~\cite[Chap.~2]{Melrose1Book} or \cite[Chap.~11]{Melrose}): by means of the {\it proper} response tensor, Eq.~\eqref{eq_fundPropRespRel}, we express the induced four-current in terms of the
total four-potential. Assuming linearity and homogeneity of the material, this yields
\begin{equation}
 j\ind^\mu(\vec x, t) = \int \! \de^3 \vec x'\! \int \! c \, \de t' \, \h \widetilde \chi\indices{^\mu_\nu}(\vec x- \vec x', t - t') \h A^\nu(\vec x', t') \,,
\end{equation}
or in Fourier space,
\begin{equation}
j\ind^\mu(\vec k, \omega)=\widetilde \chi\indices{^\mu_\nu}(\vec k, \omega) \h A^\nu(\vec k, \omega) \,.
\end{equation}
Thus, we arrive at the integro-differential equation
\begin{equation} \label{eq_FundWaveMedia_realspace}
\begin{aligned}
& \left(\eta\indices{^\mu_\nu}\mh\left(\frac{1}{c^2}\frac{\partial^2}{\partial t^2}-\Delta\right)+\frac{\partial}{\partial x_\mu}\frac{\partial}{\partial x^\nu}\right) \! A^\nu(\vec x, t) \\[3pt]
& =\mu_0\int\!\de^3 \vec x'\! \int \! c \, \de t' \,\h\widetilde \chi\indices{^\mu_\nu}(\vec x- \vec x', t - t') \h A^\nu(\vec x', t')\,,
\end{aligned}
\end{equation}
which can be written equivalently in Fourier space as
\begin{equation}\label{eq_FundWaveMedia}
\left(\left(-\frac{\omega^2}{c^2}+|\vec k|^2 \right)\!\eta\indices{^\mu_\nu}-k^\mu k_\nu-\mu_0\,\widetilde \chi\indices{^\mu_\nu}(\vec k,\omega)\right)
\!A^\nu(\vec k,\omega)=0\,.
\end{equation}
This is the general, microscopic, manifestly Lorentz-covariant wave equation for the electromagnetic four-potential in materials, which 
is well-known in plasma physics (see e.g.~\cite[Sec.~2.1.1]{Melrose}).
It depends on the concrete mate\-{}rial under consideration only through the proper response tensor $\widetilde\chi$, which is related to the fundamental response tensor $\chi$ via the Dyson type Eq.~\eqref{eq_Dyson}.

By means of the general form \eqref{generalform1} of any fundamental response tensor, we can also
write the fundamental wave equation in materials, Eq.~\eqref{eq_FundWaveMedia}, in terms of the scalar and the 
vector potential as well as the {\itshape spatial} part of the proper response tensor as follows:
\begin{align}
-\frac{c \vec k^{\rm T}}{\omega} \mh \left(\frac{\omega^2}{c^2}+\mu_0\h \tsr{\widetilde\chi}\h \right)\!\hh \vec A+
\frac{c \vec k^{\rm T}}{\omega} \mh \left(\frac{\omega^2}{c^2} + \mu_0\hh\tsr{\widetilde\chi} \h \right) \mh \frac{c\vec k}{\omega} \, \frac{\varphi}{c}=0\,, \label{discard} \\[5pt]
-\left( \h \frac{\omega^2}{c^2} \mh \left(1-\frac{c^2|\vec k|^2}{\omega^2}+\frac{c\vec k}{\omega} \h \frac{c\vec k^{\rm T}}{\omega}\right)+\mu_0 \hh \tsr{\widetilde\chi} \h \right)\!\hh\vec A+
 \left(\frac{\omega^2}{c^2}+\mu_0 \hh \tsr{\widetilde\chi}\h\right) \mh \frac{c\vec k}{\omega} \, \frac{\varphi}{c} =0\,.\label{eq_fundvecpot}
\end{align}
Note that these equations are not independent of each other: the first equation follows from the second one by multiplying through with $c\vec k^{\rm T}/\omega$, and hence Eq.~\eqref{discard} can be discarded. We remark that Eq.~\eqref{eq_fundvecpot} is still Lorentz covariant (albeit not manifestly), because it has been derived from the Lorentz-covariant equation \eqref{eq_FundWaveMedia} (see the discussion in Ref.~\cite{EDOhm}).

\subsubsection{Gauge-fixed wave equations}\label{sec_gaugefixed}

Conceptually, the problem with the above wave equations for the gauge-potential lies in the gauge freedom. This implies that the physical solutions of the wave equation are not well-defined unless we fix a gauge (cf.~\cite[Sec.~2.1]{Melrose1Book}). 
In fact, by the constraints \eqref{con_1}--\eqref{con_2} we have the equality
\begin{equation}
\widetilde \chi\indices{^\mu_\nu}(k) \, k^\nu=0\,.
\end{equation}
This implies that the pure gauges, $A^\nu(x)=\partial^\nu f(x)$, or in Fourier space,
\begin{equation} \label{pure}
A^\nu(k)={\rm i} k^\nu f(k) \,,
\end{equation}
always solve the wave equation in materials \eqref{eq_FundWaveMedia}. Furthermore, by the linearity of this equation, it follows that with any solution $A^\mu$, the gauge-trans\-{}formed four-potential $A^\mu+\partial^\mu \mh f$ also solves the wave equation in materials.

Both for the convenience of the reader and for later purposes, we will now spell out explicitly the wave equation \eqref{eq_FundWaveMedia} and its three-dimensional counterpart \eqref{eq_fundvecpot}
in fixed gauges (cf.~\cite[Sec.~2.1]{Melrose1Book}). Concretely, we consider the Lorenz gauge, the temporal gauge and the Coulomb gauge.

\bigskip \noindent
{\itshape Lorenz gauge.}---Here, the four-potential fulfills the Lorenz gauge condition,
\begin{equation}\label{eq_LorentzCond}
\partial_\mu A^\mu \h \equiv \h \frac{1}{c^2} \h \partial_t \varphi + \nabla\cdot\vec A \h =\h 0\,.
\end{equation}
For such four-potentials, the wave equation \eqref{eq_FundWaveMedia_realspace} reverts to
\begin{equation}
\bigg( \frac{1}{c^2} \h \frac{\partial^2}{\partial t^2} - \Delta \bigg) A^\mu(\vec x, t) 
-\mu_0\int\!\de^3 \vec x' \! \int \! c \, \de t' \,\h\widetilde \chi\indices{^\mu_\nu}(\vec x- \vec x', t - t') \, A^\nu(\vec x', t')=0\,,
\end{equation}
which can be written compactly as
\begin{equation} \label{compact}
(\hh\Box-\mu_0\h\widetilde \chi\h)\h A=0\,.
\end{equation}
Although this wave equation already displays a certain clarity as compared to the fundamental wave equation in materials,
it is still a tensorial equation in which the Minkowski (i.e.~$4 \times 4$) tensor $\widetilde \chi$ couples all four 
components of the gauge potential $A=(\varphi/c,\vec A)^{\rm T}$. We therefore also deduce a closed equation for the vector potential $\vec A$ in the Lorenz gauge: By Eq.~\eqref{eq_LorentzCond}, the scalar potential $\varphi$ can be expressed in terms of the vector potential $\vec A$ as
\begin{equation} \label{elimphi}
\frac{1}{c}\h\varphi(\vec k,\omega)=\frac{c\h\vec k}{\omega}\cdot\vec A(\vec k,\omega)\,.
\end{equation}
Eliminating by this the scalar potential from Eq.~\eqref{eq_fundvecpot}, we find after some manipulations
the desired closed equation for the vector potential:
\begin{equation}\label{eq_FundWaveMedia3dim}
\left( {-\frac{\omega^2}{c^2}+|\vec k|^2} - 
\mu_0\h\tsr{\widetilde \chi}(\vec k,\omega)\left(\tsr 1-\frac{c^2|\vec k|^2}{\omega^2} \, \tsr P_{\mathrm L}(\vec k)\right)\right) \mh \vec A(\vec k,\omega)=0\,.
\end{equation}
To simplify matters further, we introduce the {\it wave propagation tensor in the Lorenz gauge} as
\begin{equation}\label{eq_alphal}
\tsr\alpha_{\mathcal L}(\vec k,\omega)= \tsr{\widetilde \chi}(\vec k,\omega)\left(\tsr 1-\frac{c^2|\vec k|^2}{\omega^2} \, \tsr P_{\mathrm L}(\vec k)\right).
\end{equation}
With this, the cartesian wave equation \eqref{eq_FundWaveMedia3dim} can be written compactly as
\begin{equation}\label{eq_waveEqPropTensor}
 \left( \h -\frac{\omega^2}{c^2} + |\vec k|^2 - \mu_0 \h \tsr{\alpha}_{\mathcal L}(\vec k, \omega) \right) \! \vec A(\vec k, \omega) = 0 \,,
\end{equation}
which is in fact a well-known equation (see \cite[Eqs.~(11.3)--(11.6)]{Melrose}; note, how\-{}ever, that this book works
with the temporal gauge).

We remark that the wave-propagation tensor \eqref{eq_alphal} can be characterized as the total functional derivative (see \cite[Sec.~4.2]{ED1}) of the induced current with respect to the total vector potential in the Lorenz gauge:
\begin{equation} \label{rhs}
 \tsr \alpha_{\mathcal L} \h = \h \frac{\de \vec j_{\rm ind}}{\de \vec A_{\rm tot}} \h \equiv \h \frac{\delta \vec j_{\rm ind}}{\delta \vec A_{\rm tot}} + \frac{\delta \vec j_{\rm ind}}{\delta \varphi_{\rm tot}} \h \frac{\delta \varphi_{\rm tot}}{\delta \vec A_{\rm tot}} \,.
\end{equation}
This can again be seen by combining the general form \eqref{generalform1} of the fundamental response tensor with Eq.~\eqref{elimphi} in the Lorenz gauge. For the right hand side of Eq.~\eqref{rhs}, we thereby obtain
\begin{equation}
 \frac{\de \vec j_{\rm ind}(\vec k, \omega)}{\de \vec A_{\rm tot}(\vec k, \omega)} = \tsr{\widetilde \chi}(\vec k, \omega) - \h  \tsr {\widetilde \chi}(\vec k, \omega) \, \frac{c \vec k}{\omega} \h \frac{c \vec k^{\rm T}}{\omega} \,,
\end{equation}
which coincides with the definition of $\alpha_{\mathcal L}$\h, Eq.~\eqref{eq_alphal}. In particular, in terms of the wave propagation tensor we can write down the linear relation (cf.~\cite[Eq.~(11.3)]{Melrose})
\begin{equation}
\vec j\ind(\vec k,\omega)=\tsr\alpha_{\mathcal L}(\vec k,\omega) \h \vec A_{\rm tot}(\vec k,\omega)\,, \smallskip
\end{equation}
which does not involve the scalar potential.

\bigskip \noindent
{\itshape Temporal gauge.}---By setting $\varphi = 0$ in Eqs.~\eqref{discard}--\eqref{eq_fundvecpot}, we obtain the equations for the vector potential in the temporal gauge:
\begin{align}
-\frac{c \vec k^{\rm T}}{\omega} \mh \left(\frac{\omega^2}{c^2}+\mu_0\h \tsr{\widetilde \chi}(\vec k, \omega)\h \right)\!\hh \vec A(\vec k, \omega) =0\,, \label{temp_1} \\[5pt]
-\left( \h \frac{\omega^2}{c^2} \mh \left(1-\frac{c^2|\vec k|^2}{\omega^2}+\frac{c\vec k}{\omega} \h \frac{c\vec k^{\rm T}}{\omega}\right)+\mu_0 \hh \tsr{\widetilde \chi}(\vec k, \omega) \h \right)\!\hh\vec A(\vec k, \omega)=0\,. \label{temp_2}
\end{align}
As mentioned above, these two equations are not independent of each other. The second equation already represents a wave equation for the vector potential, which we can further transform as follows: By Eq.~\eqref{temp_1}, we have
\begin{equation}
 -\frac{\omega^2}{c^2} \h \frac{c\vec k}{\omega} \h \frac{c \vec k^{\rm T}}{\omega} \h \vec A(\vec k, \omega) = \frac{c \vec k}{\omega} \h \frac{c \vec k^{\rm T}}{\omega} \h \mu_0 \hh \tsr{\widetilde \chi}(\vec k, \omega) \h \vec A(\vec k, \omega) \,.
\end{equation}
By putting this into Eq.~\eqref{temp_2}, we obtain
\begin{equation}
 -\left( \h \frac{\omega^2}{c^2} \mh \left(1-\frac{c^2|\vec k|^2}{\omega^2}\right)+\mu_0 \left( \tsr 1 - \frac{c\vec k}{\omega} \h \frac{c \vec k^{\rm T}}{\omega} \right) \tsr{\widetilde \chi}(\vec k, \omega) \h \right)\!\hh\vec A(\vec k, \omega)=0\,,
\end{equation}
which is equivalent to
\begin{equation} \label{wave_temporal}
 \left( -\frac{\omega^2}{c^2} + |\vec k|^2 -\mu_0 \left( \tsr 1 - \frac{c^2 |\vec k|^2}{\omega^2} \h \tsr P_{\rm L}(\vec k) \right) \tsr{\widetilde \chi}(\vec k, \omega)\h \right) \vec A(\vec k, \omega) = 0 \,.
\end{equation}
This can again be written in the form 
\begin{equation}\label{eq_waveEqPropTensor_temporal}
 \left( \h -\frac{\omega^2}{c^2} + |\vec k|^2 - \mu_0 \h \tsr{\alpha}_{\mathcal T}(\vec k, \omega) \right) \! \vec A(\vec k, \omega) = 0 \,,
\end{equation}
if we define the {\itshape wave propagation tensor in the temporal gauge} as
\begin{equation} \label{eq_alphat}
 \tsr \alpha_{\mathcal T}(\vec k, \omega) = \left( \tsr 1 - \frac{c^2 |\vec k|^2}{\omega^2} \h \tsr P_{\rm L}(\vec k) \right) \tsr{\widetilde \chi}(\vec k, \omega) \,.
\end{equation}
This expression has to be compared to Eq.~\eqref{eq_alphal} in the Lorenz gauge.

\bigskip \noindent
{\itshape Coulomb gauge.}---Here, the vector potential is purely transverse, i.e.,
\begin{equation}
\vec k\cdot\vec A=0\,.
\end{equation}
Using this, one obtains from Eq.~\eqref{eq_fundvecpot} the equation
\begin{equation}\label{eq_vecpotCoulombGauge}
\left(-\frac{\omega^2}{c^2}+|\vec k|^2-\mu_0\hh\tsr{\widetilde\chi}(\vec k,\omega)\right)\!\hh\vec A
+\left(\frac{\omega^2}{c^2} + \mu_0\hh\tsr{\widetilde\chi}(\vec k,\omega)\right)\mh\frac{c\vec k}{\omega} \, \frac{\varphi}{c}=0
\end{equation}
for the coupled vector and scalar potentials $\vec A$ and $\varphi$. In contrast to the previous two cases, the scalar potential $\varphi$ cannot be eliminated from Eq.~\eqref{eq_vecpotCoulombGauge}, and hence this equation
cannot be rewritten in terms of a wave-propagation tensor.

\subsubsection{Wave equation for the electric field} \label{wave_elec}

In this subsection, we will demonstrate how the physical, i.e.~gauge-independent content can be extracted from the fundamental, gauge-dependent wave equation \eqref{eq_FundWaveMedia}. We start with the following fact from classical electrodynamics:

\bigskip \noindent
{\bfseries Lemma.} The scalar potential $\varphi$ and the longitudinal part of the vector potential $\vec A_{\rm L}$ can generally be written as
\begin{align}
\varphi(\vec x,t)&=-\partial_t f(\vec x,t)+\frac{1}{4\pi\varepsilon_0}\int \! \de^3\vec x'\,\frac{\rho(\vec x',t)}{|\vec x-\vec x'|}\,,\label{eq_scalarpot}\\[2pt]
\vec A_{\rm L}(\vec x,t)&=\nabla f(\vec x,t)\,,\label{eq_longvecpot}
\end{align}
or in Fourier space,
\begin{align}
 \varphi(\vec k, \omega) & = \j\omega f(\vec k, \omega) + \frac{\rho(\vec k, \omega)}{\varepsilon_0 |\vec k|^2} \,, \label{eq_scalarpot_FT} \\[2pt]
 \vec A_{\rm L}(\vec k, \omega) & = \j \vec k \h f(\vec k, \omega) \,, \label{eq_longvecpot_FT}
\end{align}
where $f$ is a suitable scalar function. This means, the scalar potential is generally given by the sum of the Coulomb potential and a pure gauge, while the longitudinal vector potential is always given by a pure gauge.

\pagebreak \noindent
{\bfseries Proof.} We first consider the situation in the Coulomb gauge, $\nabla\cdot\vec A=0$, where the longitudinal part of the vector potential vanishes identically. The longitudinal part of the electric field is then given by
\begin{equation}
 \vec E_{\mathrm L}(\vec x, t) = -\nabla\varphi(\vec x, t)\,,
\end{equation}
and from Gauss' law \eqref{eq_Maxfund1} it follows that
\begin{equation}
-\Delta\varphi(\vec x,t)=\frac{\rho(\vec x,t)}{\varepsilon_0}\,.
\end{equation}
This is the Poisson equation, whose solution can be expressed as
\begin{equation} \label{poisson_sol}
 \varphi(\vec x, t) = \int \! \de^3 \vec x'\! \int \! c \, \de t' \, \h v(\vec x - \vec x', \h t - t') \, \rho(\vec x', t') \,.
\end{equation}
Here, $v$ denotes the Coulomb interaction kernel,
\begin{equation}
 v(\vec x - \vec x', \h t - t') = \frac{1}{4 \pi \varepsilon_0} \, \frac{\delta(c \h t - c \h t')}{|\vec x - \vec x'|} \,,
\end{equation}
whose Fourier transform is given by
\begin{equation} \label{cint_FT}
 v(\vec k) \equiv v(\vec k, \omega) = \frac{1}{\varepsilon_0 |\vec k|^2} \,.
\end{equation}
We conclude that in the Coulomb gauge, the scalar potential is given by the Coulomb potential, while the longitudinal vector potential vanishes. In particular, Eqs.~\eqref{eq_scalarpot}--\eqref{eq_longvecpot} are fulfilled with $f(\vec x, t) \equiv 0$. 

Now, the general solution of the equation of motion for the four-potential, Eq.~\eqref{fund_eom}, 
can be obtained by a gauge transformation from a special solution in a fixed gauge. Concretely, any four-potential $A^\mu$ can be represented in terms of the four-potential $A^\mu_{\mathcal C}$ in the Coulomb gauge by
\begin{equation}
 A^\mu = A_{\mathcal C}^\mu + \partial^\mu \mh f \,,
\end{equation}
where $f$ is an arbitrary scalar function, and $\partial^\mu \mh f$ is called a pure gauge. Combining this equation with the above results for the Coulomb gauge yields 
immediately Eqs.~\eqref{eq_scalarpot}--\eqref{eq_longvecpot} in real space, or Eqs.~\eqref{eq_scalarpot_FT}--\eqref{eq_longvecpot_FT} in Fourier space. In particular, the freedom of choosing  the  function $f$ precisely corresponds to the gauge freedom.  \qed

\bigskip
The upshot of our Lemma is that we can control the gauge arbitrariness explicitly in the form of the arbitrary scalar function~$f$.
With this, we now come back to the fundamental wave equation \eqref{eq_FundWaveMedia}, or its three-dimensional counterpart \eqref{eq_fundvecpot}. We first decompose the vector potential 
into its longitudinal and transverse parts, $\vec A=\vec A_{\mathrm L}+\vec A_{\mathrm T}$, whereby we obtain
\begin{equation}
 \left( - \frac{\omega^2}{c^2} + |\vec k|^2 - \mu_0 \hh \tsr{\widetilde\chi} \h \right)\!\hh \vec A_{\rm T} +
 \left(\frac{\omega^2}{c^2}+\mu_0 \hh \tsr{\widetilde\chi}\h\right) \left( \frac{c\vec k}{\omega} \, \frac{\varphi}{c} - \vec A_{\rm L} \right) =0\,.
\end{equation}
By plugging in the representation \eqref{eq_scalarpot_FT}--\eqref{eq_longvecpot_FT}, we then see that all contributions involving the pure gauge
{\itshape cancel exactly}, and thus we finally~arrive~at
\begin{equation} \label{compare_1}
\begin{aligned}
& \left(-\frac{\omega^2}{c^2}+|\vec k|^2-\mu_0\hh\tsr{\widetilde\chi}(\vec k,\omega)\right)\!\hh\vec A_{\mathrm T}(\vec k, \omega) \\[3pt]
& \h + \h \left(\frac{\omega^2}{c^2}+\mu_0\hh\tsr{\widetilde\chi}(\vec k,\omega)\right)\mh\frac{\vec k}{\omega} \, v(\vec k) \h \rho(\vec k,\omega)=0\,,
\end{aligned}
\end{equation}
where $v(\vec k)$ denotes the Coulomb interaction kernel (see Eq.~\eqref{cint_FT}).
In other words, the gauge-dependent parts drop out, giving rise to a {\itshape gauge-indepen\-{}dent} equation for the transverse part of the vector potential and the charge density. Interestingly, this gauge-independent equation formally coincides with 
its counterpart in the Coulomb gauge, Eq.~\eqref{eq_vecpotCoulombGauge}, if we resubstitute the charge density in terms of the scalar potential by means of Eq.~\eqref{poisson_sol}.

We will now further reformulate the gauge-independent equation \eqref{compare_1} in terms of the electric field. For this purpose, we  multiply this equation through 
by \h$\j\omega$ and use the gauge-independent relations
\begin{align}
\vec E_{\mathrm L}(\vec k,\omega)&=-\j\vec k\,v(\vec k) \h \rho(\vec k,\omega)\,,\\[5pt]
\vec E_{\mathrm T}(\vec k,\omega)&=\j\omega \vec A_{\mathrm T}(\vec k,\omega) \,,
\end{align}
which follow from Eq.~\eqref{eq_EAphi}
by projecting onto the longitudinal\h/\h transverse parts and using Eqs.~\eqref{eq_scalarpot}--\eqref{eq_longvecpot}.
This leads to the equation
\begin{equation}\label{eq_fundWaveEqElField}
\left(-\frac{\omega^2}{c^2}+|\vec k|^2-\mu_0\hh\tsr{\widetilde\chi}(\vec k,\omega)\right)\mh\vec E_{\mathrm T}(\vec k,\omega)+
\left(-\frac{\omega^2}{c^2}-\mu_0\hh\tsr{\widetilde\chi}(\vec k,\omega)\right)\mh\vec E_{\mathrm L}(\vec k,\omega)=0\,,
\end{equation}
which is equivalent to
\begin{equation}\label{eq_masterformula}
\left(-\frac{\omega^2}{c^2}+|\vec k|^2-\mu_0\tsr{\widetilde\chi}(\vec k,\omega)\right)\mh\vec E(\vec k,\omega)-|\vec k|^2\h \vec E_{\mathrm L}(\vec k,\omega)=0\,.
\end{equation}
This is not yet our final result for the wave equation of the electric field in media.
In order to get rid of the last term with the longitudinal electric field, we employ once more Amp\`{e}re's law \eqref{eq_Maxfund4}, which implies the identity\,\footnote{The same identity also follows from the continuity equation and  Gauss' law \eqref{eq_Maxfund1}:
\begin{equation*}
 \vec j_{\rm L}(\vec k, \omega) \h = \h \frac{\vec k \h (\vec k \cdot \vec j(\vec k, \omega))}{|\vec k|^2} \h = \h \frac{\omega}{|\vec k|^2} \, \vec k \h \rho(\vec k, \omega)\h  =\h  \j\omega \hh \varepsilon_0 \h \vec E_{\rm L}(\vec k, \omega) \,.
\end{equation*}
This is similar to the remark in \cite[Eq.~(4.20)]{ED1}.}
\begin{equation} \label{id_1}
 \vec E_{\mathrm L}(\vec k, \omega) = \frac{1}{\j \omega \h \varepsilon_0 } \, \vec j_{\rm L}(\vec k, \omega) \smallskip
\end{equation}
between the longitudinal part of the electric field and the spatial current. In the absence of external sources, $\vec j \hh \equiv \hh \vec j_{\rm tot} = \vec j_{\rm ind}$, we can then use Ohm's law in the form
\begin{equation} \label{ohm}
 \vec j(\vec k, \omega) = \tsr{\widetilde \sigma}(\vec k, \omega) \h \vec E(\vec k, \omega) = \frac{1}{\j\omega} \h \tsr{\widetilde \chi}(\vec k, \omega) \h \vec E(\vec k, \omega) 
\end{equation}
to eliminate the current in terms of the total electric field. Thus, we obtain the relation \smallskip
\begin{equation}
 \vec E_{\mathrm L}(\vec k, \omega) = -\frac{1}{\varepsilon_0 \h \omega^2} \h \tsr P_{\mathrm L}(\vec k) \h \tsr{\widetilde \chi}(\vec k, \omega) \h \vec E(\vec k, \omega) \,. \smallskip
\end{equation}
By plugging this into Eq.~\eqref{eq_masterformula}, we finally arrive at
\begin{equation} \label{eq_equivalent_masterformula}
 \left( -\frac{\omega^2}{c^2} + |\vec k|^2 -\mu_0\left( \tsr 1 - \frac{c^2 |\vec k|^2}{\omega^2} \h \tsr P_{\mathrm L} (\vec k) \right) \tsr{\widetilde \chi}(\vec k, \omega) \right) \mh \vec E(\vec k, \omega) = 0 \,.
\end{equation}
This is the fundamental wave equation for the electric field in materials. At the same time, it is the gauge-independent wave equation in materials. In terms of the wave propagation tensor in the temporal gauge, Eq.~\eqref{eq_alphat}, the wave equation for the electric field can be written compactly as
\begin{equation} \label{eq_compact_masterformula}
  \left( \h -\frac{\omega^2}{c^2} + |\vec k|^2 - \mu_0 \h \tsr{\alpha}_{\mathcal T}(\vec k, \omega) \right) \! \vec E(\vec k, \omega) = 0 \,.
\end{equation}
Indeed, this equation has the same form as the wave equation \eqref{eq_waveEqPropTensor_temporal} for the vector potential in the temporal gauge, and can hence be rederived from the latter by using the relation \h$\vec E = \j\omega \vec A$ \h(which holds in the temporal gauge).

\subsubsection{Connection to the dielectric tensor} \label{conn_diel}

In this subsection, we put our wave equations into perspective by deriving from them an even simpler condition for the electromagnetic wave propagation in materials.
The full electromagnetic Green function, Eq.~\eqref{full_GF}, can be interpreted as an integral kernel relating the total fields 
to the external sources as (see e.g.~\cite[Eq.~(2.1.4)]{Melrose1Book})
\begin{equation}\label{eq_deffullGF}
A^\mu(x)=\int \! \de^4 x'\,D\indices{^\mu_\nu}(x - x') \, j^\nu\ext(x')\,,
\end{equation}
or in Fourier space,
\begin{equation} \label{eq_deffullGF_Fourier}
 A^\mu(\vec k, \omega) = D\indices{^\mu_\nu}(\vec k, \omega) \, j^\nu\ext(\vec k, \omega) \,.
\end{equation}
Na\"{i}vely, one might conclude from this equation that in the absence of external sources the total four-potential is alway zero.
However, by inversion of the full electromagnetic Green function, Eq.~\eqref{eq_deffullGF_Fourier} yields the formal (see \cite[Sec.~3.3]{ED1} and \cite{EDWave}) {\it inhomogeneous wave equation}
\begin{equation}
D^{-1}(\vec k,\omega) \h A(\vec k,\omega)=j\ext(\vec k,\omega) \,. \smallskip
\end{equation}
Now, if the external four-current vanishes, the total four-potential can still be non-zero if it fulfills the corresponding 
{\it homogeneous wave equation} (see \cite[Chap.~2]{Melrose1Book})
\begin{equation}
D^{-1}(\vec k,\omega) \h A(\vec k,\omega)=0\,,\label{eq_condition} \smallskip \vspace{2pt}
\end{equation}
i.e.,~if the four-potential lies in the kernel of the inverse Green function. On the other hand, the Schwinger--Dyson equation \eqref{eq_Dyson1} between the full and the free electromagnetic Green function formally (see \cite{EDWave}) implies that
\begin{equation}
D^{-1}=D_0^{-1}-\widetilde\chi\,.
\end{equation}
Plugging in the explicit expression \cite[Sec.~3.3]{ED1}
\begin{equation} \label{invert}
 (D_0^{-1})\indices{^\mu_\nu} = \frac{1}{\mu_0} \h (\eta\indices{^\mu_\nu} \h \Box + \partial^\mu \partial_\nu) \,,
\end{equation}
it follows immediately that the condition \eqref{eq_condition} is equivalent
to the fundamental wave equation for the four-potential, Eq.~\eqref{eq_FundWaveMedia_realspace}. 
The general interpretation of the wave equation in materials can therefore easily be stated: the field lies in the kernel
of the inverse of a response function, which expresses total in terms of external quantities.
This conclusion can be more spectacularly generalized by the following theorem, which constitutes the main result of this article: 

\bigskip
\noindent
{\bfseries Theorem.} The Lorentz-covariant, four-dimensional wave equation in media given by Eq.~\eqref{eq_FundWaveMedia}, i.e.,
\begin{equation}
\left(\left(-\frac{\omega^2}{c^2}+|\vec k|^2 \right)\eta\indices{^\mu_\nu}-k^\mu k_\nu-\mu_0\,\widetilde \chi\indices{^\mu_\nu}(\vec k,\omega)\right)
\!A^\nu(\vec k,\omega)=0\,,
\end{equation}
is equivalent to the simple condition
\begin{equation} \label{simple_cond}
\tsr\varepsilon_{\rm r}(\vec k,\omega) \h \vec E(\vec k,\omega)=0\,,
\end{equation}
meaning that the electric field component of the wave in the medium lies in the kernel (null-space) of the dielectric tensor.

\bigskip \noindent
{\bfseries Proof.} We have already shown that the fundamental, Lorentz-covariant wave equation \eqref{eq_FundWaveMedia} is equivalent to the three-dimensional wave equation for the electric field, Eq.~\eqref{eq_equivalent_masterformula}. By factoring out $(-\omega^2/c^2 + |\vec k|^2)$, the latter is equivalent to
\begin{equation}
 \left( \h \tsr 1 - \mathbbmsl D_0(\vec k, \omega) \left( \tsr 1 - \frac{c^2 |\vec k|^2}{\omega^2} \h \tsr P_{\rm L}(\vec k) \right) \tsr{\widetilde \chi}(\vec k, \omega) \right) \mh \vec E(\vec k, \omega) = 0 \,,
\end{equation}
with $\mathbbmsl D_0$ given by Eq.~\eqref{gf}. In terms of the electric solution generator \eqref{eq_OpE}, this equation can be further simplified to
\begin{equation} \label{coml_prf}
 \left( \h \tsr 1 + \frac{1}{\varepsilon_0 \h \omega^2} \h \tsr{\mathbbmsl E}(\vec k, \omega) \h \tsr{\widetilde \chi}(\vec k, \omega) \right) \mh \vec E(\vec k, \omega) = 0 \,.
\end{equation}
By the universal relations \eqref{eq_standardRel1} and \eqref{cond_rel_2}, the term in brackets equals pre-\linebreak cisely the dielectric tensor, and this completes the proof. \qed

\bigskip
We particularly stress that neither Eq.~\eqref{eq_FundWaveMedia} nor \eqref{simple_cond} are actually new. In fact, both are already well-established.
While the fundamental, covariant wave equation is, as mentioned above, well-known in plasma physics \cite{Melrose1Book, Melrose}, the wave equation in terms of the dielectric tensor can be found, for example, in \cite[pp.~232~f.]{Dolgov}. These equations do therefore {\it not} constitute new hypo\-{}theses introduced 
by the authors of this article. Instead, we only establish their equivalence, which will prove crucial for the discussion of the refractive index later on.

We conclude this subsection with two observations that are made after rewriting the wave equation for the electric field in the form \eqref{simple_cond}. Firstly, an alternative derivation of the wave equation for the electric field can be given by combining Eq.~\eqref{eq_tsrE} for the total electric field, $\vec E \equiv \vec E_{\rm tot}$\h, \linebreak with Ohm's law in the form \eqref{eq_properOhmLaw}: using that in the absence of external sources, $\vec j \equiv \vec j_{\rm tot} = \vec j_{\rm ind}$\h, we obtain
\begin{equation}
 \vec E(\vec k, \omega) = \frac{1}{\j \omega \h \varepsilon_0} \, \tsr{\mathbbmsl E}(\vec k, \omega) \, \vec j(\vec k, \omega) = \frac{1}{\j\omega \h \varepsilon_0} \, \tsr{\mathbbmsl E}(\vec k, \omega) \h \tsr{\widetilde \sigma}(\vec k, \omega) \h \vec E(\vec k, \omega) \,,
\end{equation}
which by the universal relation \eqref{cond_rel_2} is equivalent to Eq.~\eqref{simple_cond}. On the other hand, a closed wave equation for the magnetic field in materials cannot be obtained in general, 
because neither the current density nor the electric field can be expressed entirely in terms of the magnetic field. It is only in the isotropic limit that the wave equations for the longitudinal and transverse electric field components decouple, and the transverse electric field can be expressed entirely in terms of the magnetic field (see Sec.~\ref{sec_iso}).

Secondly, the equation \eqref{simple_cond} makes it very clear why the Standard Approach to the wave equation in materials fails:
As discussed in Sec.~\ref{subsubsec:naive}, there one re-expresses Amp\`{e}re's law for the external fields
in terms of the total fields by means of the dielectric function and the inverse magnetic permeability. In other words, one eliminates
the external fields in favor of the total fields using the relations $\vec D=\varepsilon_0 \h \varepsilon_{\rm r} \hh \vec E$ and $\vec H=\mu_0^{-1}\mu_{\rm r}^{-1}\vec B$.
This, however, is not possible, because in the case of light waves in media, total fields lie in the kernels of the respective inverse response functions 
(within the limits of the idealization used in the derivation of the refractive index, see also the remark at the end of Sec.~\ref{sec_iso}).

\subsection{Speed of light in materials}\label{subsec_speed_light}

We now come back to our original objective, the speed of light in media. This notion still remains to be defined in the microscopic framework.
For that purpose, we investigate the general form of the solution to the equation \eqref{simple_cond} and use once more
a standard procedure of microscopic approaches to electrodynamics of materials: the speed of light is {\it defined}
from the dispersion relation in materials, and the latter is in turn obtained from setting the wave-operator to zero (see e.g.~\cite[Chap.~2]{Melrose1Book} and \cite[Chap.~2]{Platzmann}). Concretely, for $\vec k$ and $\omega$ being fixed, Eq.~\eqref{simple_cond} is an ordinary homogeneous linear equation, which
implies that the Fourier amplitude $\vec E(\vec k,\omega)$ is zero if the matrix of the dielectric tensor is invertible.
Hence, for the amplitude to be non-trivial we find the condition
\begin{equation} \label{eq_defdispersion}
\det\tsr\varepsilon_{\rm r}(\vec k,\omega) = 0 \,, \smallskip
\end{equation}
which is referred to as the {\itshape dispersion equation}. By Eq.~\eqref{eq_compact_masterformula}, this is equivalent to the well-known equation (see e.g.~\cite[Chap.~11]{Melrose}, \cite[Eq.~(2.14)]{Piel} and \cite[Eq.~(5.11)]{Fitzpatrick}) 
\begin{equation} \label{eq_defdispersion_equivalent} 
\det\left(\left(-\frac{\omega^2}{c^2}+|\vec k|^2\right)\tsr 1-\mu_0 \h \tsr{\alpha}_{\mathcal T}(\vec k,\omega)\right) = 0\,.
\end{equation}
For each wave-vector $\vec k$, this is an implicit equation for the determination of the frequency $\omega$, thereby giving the {\itshape dispersion relation} in the material,
\begin{equation}
\omega=\omega_{\vec k \lambda} \,.\label{eq_dispersion}
\end{equation}
Here, we have introduced the index $\lambda$ to indicate that in materials, the 
dis\-{}persion equation \eqref{eq_defdispersion} may have several solutions for each $\vec k$.
In particular, we note that for \h$\alpha_{\mathcal T}(\vec k, \omega) \equiv 0$, one recovers the vacuum dispersion relation $\omega_{\vec k}=c|\vec k|$ (i.e. the dispersion relation of free light waves).

We are now in a position to define the speed of light in materials: Recall that the dispersion relation of free light waves in vacuo reads
\begin{equation}
\omega_{\vec k} =c \h |\vec k|\,. 
\end{equation}
Correspondingly, we {\it define} the speed of light in materials, $u = u_{\vec k \lambda}$\h, 
by simply factoring out $|\vec k|$ in the dispersion relation \eqref{eq_dispersion}, such that
\begin{equation} \label{def_u}
\omega_{\vec k\lambda} =: u_{\vec k\lambda} \h |\vec k| \,.
\end{equation}
In other words, the wavelength-dependent speed of light in materials is defined as the phase velocity of the corresponding oscillation mode.
Correspondingly, the wavelength dependent index of refraction is then given by \cite[Sec.~2.2.4]{Melrose1Book}
\begin{equation}
n_{\vec k\lambda}=\frac{c}{u_{\vec k\lambda}} = \frac{c|\vec k|}{\omega_{\vec k\lambda}} \,. \label{eq_defRI} \smallskip
\end{equation}
We remark that this redefinition of the speed of light in media (previously defined by the prefactor in the standard wave equation \eqref{original_definition}) through the dispersion relation is in fact standard in plasma physics (see \cite[Sec.~2.2.4]{Melrose1Book}, \cite[p.~26]{Piel}, \cite[Eq.~(5.40)]{Fitzpatrick} and \cite[p.~107]{Kamenetskii};  cf.~also \cite[Sec.~11.2]{Melrose} and \cite[Eq.~(33.7)]{Fliessbach}).

\subsection{Isotropic limit} \label{sec_iso}

In this subsection, we come back to the isotropic limit, an idealization which is particularly important for practical applications. In this limit, the gauge-independent wave equation in materials, Eq.~\eqref{eq_equivalent_masterformula}, yields the following decoupled equations 
for the longitudinal component $\vec E_{\rm L}$ and the transverse component $\vec E_{\rm T}$ of the electric field:
\begin{align}
 \left( -\frac{\omega^2}{c^2} + |\vec k|^2 - \mu_0 \left( 1 - \frac{c^2 |\vec k|^2}{\omega^2} \right) \widetilde \chi_{\rm L}(\vec k, \omega) \right) \vec E_{\rm L}(\vec k, \omega) & = 0  \label{eq_long} \,, \\[3pt]
 \left( -\frac{\omega^2}{c^2} + |\vec k|^2 - \mu_0 \h \widetilde \chi_{\rm T}(\vec k, \omega) \right) \vec E_{\rm T}(\vec k, \omega) & = 0 \,. \label{eq_FundWaveMedia3dimSimpEfield1}
\end{align}
By our Theorem, these equations are equivalent to
\begin{align}
 \varepsilon_{\rm r, \hh L}(\vec k, \omega) \h \vec E_{\mathrm L}(\vec k, \omega) & = 0 \,, \label{iso_wave_eq_1} \\[5pt]
 \varepsilon_{\rm r, T}(\vec k, \omega) \h \vec E_{\mathrm T}(\vec k, \omega) & = 0 \label{iso_wave_eq_2} \,,
\end{align}
where the longitudinal and transverse dielectric response functions are defined by the equality
\begin{equation}
 \tsr \varepsilon_{\rm r}(\vec k, \omega) = \varepsilon_{\rm r, \hh L}(\vec k, \omega) \h \tsr P_{\rm L}(\vec k) + \varepsilon_{\rm r, T}(\vec k, \omega) \h \tsr P_{\rm T}(\vec k) \,. \smallskip
\end{equation}
In particular, the equivalence of Eq.~\eqref{eq_long} and Eq.~\eqref{iso_wave_eq_1} can be verified directly as follows: By factoring out $(-\omega^2/c^2 + |\vec k|^2)$, Eq.~\eqref{eq_long} turns into
\begin{equation} \label{zw_1}
 \left( 1 + \frac{1}{\varepsilon_0 \h \omega^2} \, \widetilde \chi_{\rm L}(\vec k, \omega) \right) \mh \vec E_{\rm L} (\vec k, \omega) = 0 \,.
\end{equation}
In terms of the proper density response function \cite[Sec.~7.1]{ED1}
\begin{equation}
 \widetilde\upchi = \frac{\delta \rho_{\rm ind}}{\delta \varphi_{\rm tot}} = \frac{1}{c^2} \h \widetilde\chi\indices{^0_0}\,, 
\end{equation}
which is related to the longitudinal current response function by \cite{ED1, Giuliani}
\begin{equation}
 \widetilde \upchi(\vec k, \omega) = -\frac{|\vec k|^2}{\omega^2} \h \widetilde \chi_{\rm L}(\vec k, \omega) \,,
\end{equation}
Eq.~\eqref{zw_1} can be written equivalently as
\begin{equation}
(1-v(\vec k)\h\hh \widetilde\upchi(\vec k,\omega)) \h \vec E_{\mathrm L}(\vec k,\omega)=0\,,
\end{equation}
where $v(\vec k)$ is the Coulomb interaction kernel (see Eq.~\eqref{cint_FT}). Hence, with 
the standard relation \cite[Eq.~(5.21)]{Giuliani}
\begin{equation}
\varepsilon_{\rm r}(\vec k,\omega)=1-v(\vec k) \h\hh \widetilde\upchi(\vec k,\omega)\,,
\end{equation}
which is generally valid in the isotropic limit under the identification of $\varepsilon_{\rm r}$ with the {\itshape longitudinal} dielectric function, we ultimately retrieve Eq.~\eqref{iso_wave_eq_1}.

We now come to the interpretation of the equations \eqref{iso_wave_eq_1}--\eqref{iso_wave_eq_2}: 
Their meaning is that in the isotropic limit, the longitudinal and transverse proper oscillations of the medium decouple. 
The respective dispersion relations are determined by (cf.~\cite[Eq.~(2.34)]{Dolgov})
\begin{align}
 \varepsilon_{\rm r, \hh L}\hh(\vec k, \h \omega_{\vec k \hh {\rm L}}\hh) &= 0 \label{plasmon} \,,\\[5pt]
 \varepsilon_{\rm r, T}\hh(\vec k, \h \omega_{\vec k {\rm T}}\hh) &= 0 \,. \label{trl}
\end{align}
Consequently, Eq.~\eqref{iso_wave_eq_1} describes plasmons, and the roots of Eq.~\eqref{plasmon} are the plasmon frequencies (see e.g.~\cite[Eq.~(4.92)]{MartinRothen} and \cite[Eq.~(18.25)]{Bechstedt}). By contrast, Eq.~\eqref{iso_wave_eq_2} describes transverse light waves, and the dispersion relation of Eq.~\eqref{trl} determines the refractive index. Thus,  the theory of plasmons
combines with the theory of transverse electromagnetic waves in media into one unified wave equation in materials given by Eq.~\eqref{eq_equivalent_masterformula} or \eqref{simple_cond} (cf.~\cite[Sec.~18.3.2]{Bechstedt}).

Finally, for the sake of completeness, we show that the transverse wave equation \eqref{iso_wave_eq_2} can be reformulated in terms of the magnetic field: By means of Faraday's law, which implies
\begin{equation} \label{far}
 \vec E_{\rm T}(\vec k, \omega) = -\frac{\omega}{|\vec k|^2} \h \vec k \times\vec B(\vec k, \omega) \,,
\end{equation}
and by using the relation \eqref{idfar}, we can write Eq.~\eqref{iso_wave_eq_2} equivalently as
\begin{equation} \label{this_1}
 -\frac{\omega}{|\vec k|^2} \, \vec k \times \left( \h \mu_{\rm r}^{-1}(\vec k, \omega) \h \vec B(\vec k, \omega)\right) = 0 \,.
\end{equation}
The magnetic field, and consequently the whole term in brackets, is a purely transverse vector field. Therefore, Eq.~\eqref{this_1} is equivalent to
\begin{equation}
 \mu_{\rm r}^{-1}(\vec k, \omega) \h \vec B(\vec k, \omega) = 0 \,.
\end{equation}
This shows that in the isotropic limit, the total magnetic field also lies in the kernel of the inverse of a response function, $\mu_{\rm r}$, which expresses total in terms of external quantities.

\subsection{Maxwell relation reconsidered} \label{recon}

We now re-investigate the Maxwell relation, $n^2=\varepsilon_{\rm r}$\h, in the light of our findings. 
In particular, we consider our results in the isotropic limit as described in the previous subsection. 
Assuming the fields to be purely transverse---which is indeed the case for light waves---the wave equation in materials is now simply given by Eq.~\eqref{eq_FundWaveMedia3dimSimpEfield1}. 
This equation is of course at variance with the standard wave equation in media, Eq.~\eqref{eq_freeWaveMediaFourierEq1}, which we have refuted already in Sec.~\ref{sec_refute}. 
We will instead compare our Eq.~\eqref{eq_FundWaveMedia3dimSimpEfield1} to the standard approximation given by Eq.~\eqref{eq_certainJust1}, i.e.~more precisely,
\begin{equation}
\left(-\frac{\omega^2}{c^2} \h \varepsilon_{\rm r, \hh L}(\vec k,\omega)+|\vec k|^2\right) \! \vec E_{\mathrm T}(\vec k,\omega) = 0\,. \label{standard_wave_eq}
\end{equation}
Note that in this equation, the optical properties are controlled by the {\it longitudinal} dielectric function, as it is usually assumed in ab initio electronic structure physics (see e.g.~\cite[App.~E]{Martin}, \cite[Sec.~2.7]{SchafWegener} or \cite[Chap.~6]{Cardona}).
As we are now going to show, this equation can be justified from Eq.~\eqref{eq_FundWaveMedia3dimSimpEfield1} at optical wavelengths.
In fact, using again the connection \eqref{eq_standardRel1} between the current response tensor and the conductivity tensor,
we reformulate the wave equation \eqref{eq_FundWaveMedia3dimSimpEfield1} as
\begin{equation}\label{eq_desired}
 \left( -\frac{\omega^2}{c^2} \left( 1 - \frac{1}{\j \omega \h \varepsilon_0} \, \widetilde \sigma_{\mathrm T}(\vec k, \omega) \right) + |\vec k|^2 \right) \mh \vec E_{\mathrm T}(\vec k, \omega) 
 = 0 \,.
\end{equation}
For optical wavelengths, we further assume that
\begin{equation} \label{lequalt}
 \widetilde\sigma_{\mathrm L}(\vec k, \omega) \h \approx \h \widetilde\sigma_{\mathrm T}(\vec k, \omega) \,,
\end{equation}
such that Eq.~\eqref{wn} implies
\begin{equation} \label{questionable}
\varepsilon_{\rm r, \hh L}(\vec k, \omega) = 1 - \frac{1}{\j \omega \h \varepsilon_0} \, \widetilde \sigma_{\mathrm L}(\vec k, \omega) \h \approx \h 1 - \frac{1}{\j \omega \h \varepsilon_0} \, \widetilde \sigma_{\mathrm T}(\vec k, \omega) \,.
\end{equation}
Substituting this into Eq.~\eqref{eq_desired} yields the desired wave equation \eqref{standard_wave_eq} for the electric field.
More generally, under the assumption \eqref{lequalt} one shows directly that
\begin{equation}
\left(-\frac{\omega^2}{c^2}+|\vec k|^2\right) \mh \varepsilon_{\rm r,T}(\vec k,\omega)=\left(-\frac{\omega^2}{c^2}\, \varepsilon_{\rm r, \hh L}(\vec k,\omega)+|\vec k|^2\right), 
\end{equation}
and with this the desired standard wave equation of ab initio electronic structure physics, Eq.~\eqref{standard_wave_eq}, follows directly from Eq.~\eqref{iso_wave_eq_2}. Similarly, the corresponding Eq.~\eqref{eq_certainJust2} for the magnetic field can be shown from Eq.~\eqref{standard_wave_eq} by converting the 
electric field into a magnetic field using Faraday's law.

Now, if we further {\itshape neglect} the wave-vector dependence of the dielectric function,
then we find from Eq.~\eqref{standard_wave_eq} the dispersion equation
\begin{equation}\label{eq_dispersionEq}
\frac{\omega_{\vec k}^2}{c^2} \, \varepsilon_{\rm r, \hh L}\hh(\omega_{\vec k}) \h = \h |\vec k|^2 \,,
\end{equation}
which is in fact common knowledge (see e.g.~\cite[Eq.~(18.24)]{Bechstedt}, or \cite[Eq. (8.33)]{Strinati}).
We remark, however, that this equation is in general not equivalent to the na\"{i}ve dispersion relation
\begin{equation} \label{disp_rel}
\omega_{\vec k}=\frac{c}{\sqrt{\varepsilon_{\rm r, \hh L}}} \, |\vec k|\,,
\end{equation}
which by Eq.~\eqref{def_u} would be equivalent to the Maxwell relation,
\begin{equation}
u = \frac{c}{\sqrt{\varepsilon_{\rm r, \hh L}}}\,.
\end{equation}
The last formula would only be correct if also the frequency dependence of the dielectric function in Eq.~\eqref{eq_dispersionEq} could be neglected 
(which may be the case for some, but not all gases \cite{Dunn, Boggs, Essen, May}). In general, the {\it dispersion equation} \eqref{eq_dispersionEq} is only 
an {\it implicit} equation for the determination of the {\it dispersion relation} \h$\omega=\omega_{\vec k}$\hh. Generally, one can show that under the assumption \eqref{lequalt} the refractive index (as defined by Eq.~\eqref{eq_defRI})
is given by the implicit equation
\begin{equation}
n^2_{\vec k}=\varepsilon_{\rm r, \hh L}(\vec k, \h \omega_{\vec k {\rm T}})\,,
\end{equation}
where $\omega_{\vec k {\rm T}}$ is defined by the condition $\varepsilon_{\rm r, T}(\vec k, \h \omega_{\vec k {\rm T}})=0$.
In summary, we have shown that the textbook wave equation \eqref{standard_wave_eq} and 
the ensuing dispersion equation~\eqref{eq_dispersionEq} can be further upheld as long as their validity is restricted to optical wavelengths.

\section{Conclusion}

Based on modern microscopic approaches to electrodynamics of media---as they are common practice in ab initio materials physics 
and axiomatized by the Functional Approach---we have subjected the standard formula for the refractive index, $n^2 = \varepsilon_{\rm r} \h \mu_{\rm r}$\h, 
to a thorough re-investigation, whereupon we \linebreak have found its untenability.
In particular, we have refuted the standard wave \linebreak equations \eqref{eq_freeWaveMediaFourierEq1}--\eqref{eq_freeWaveMediaFourierEq2}.
We have subsequently given a fully relativistic re\-{}derivation of the wave equation in materials, starting from the Lorentz-cova\-{}riant wave equation \eqref{eq_FundWaveMedia}, which is standard in plasma physics. From this, we have rederived the three-dimensional wave equations for the vector potential and the electric field (Secs.~\ref{sec_gaugefixed}--\ref{wave_elec}), 
and thereby clarified their relation to the fundamental response tensor. Furthermore, from the gauge-independent wave equation \eqref{eq_equivalent_masterformula},
we have shown that the textbook wave equations \eqref{eq_certainJust1}--\eqref{eq_certainJust2}
and correspondingly also the Maxwell-relation, $n^2=\varepsilon_{\rm r}$\h, can be justified at optical wavelengths. 
Independently of this limit though, we have shown that all wave equations can be condensed into the simple
formula \eqref{simple_cond}, which restricts the electric field in media to the kernel of the microscopic dielectric tensor.
This allows in particular for a unification of the theory of light propagation in materials and the theory of plasmons (Sec.~\ref{sec_iso}). 
Thus, this work contributes to the modern microscopic approaches to electrodynamics of materials, 
and it may represent a further step towards the ultimate goal of calculating all optical materials properties from first principles. 

\section*{Acknowledgements}
This research was supported by the Austrian Science Fund (FWF) within the SFB ViCoM, Grant No. F41, 
and by the DFG Reseach Unit FOR 723. R.\,S.~thanks the Institute for Theoretical Physics at the TU Bergakademie Freiberg for its hospitality.

\bibliographystyle{model1-num-names}
\bibliography{/net/home/lxtsfs1/tpc/schober/Ronald/masterbib}

\end{document}